\documentclass{jfm}
\usepackage{ifthen}
\usepackage{wrapfig}
\usepackage{graphicx}
\usepackage{newtxtext}
\usepackage{newtxmath}
\usepackage{natbib}
\usepackage{siunitx}
\usepackage[normalem]{ulem}
\usepackage[dvipsnames]{xcolor}
\usepackage{bm}
\usepackage{extarrows}
\usepackage{caption}
\usepackage{subcaption}
\usepackage{import}
\usepackage{pdflscape}
\usepackage{lipsum}
\usepackage{pgfplots}
\usepackage{amsmath, amssymb, amsfonts}
\usepackage{rotating}
\usepackage{adjustbox}

\pgfplotsset{compat=1.18}

\newcommand{\be}{\begin{equation}}
\newcommand{\ee}{\end{equation}} 
\newcommand{\bi}{\begin{itemize}}
\newcommand{\ei}{\end{itemize}}

\newcommand{\yaw}{\chi_{\rm yaw}}
\newcommand{\pitch}{\chi_{\rm pitch}}
\newcommand{\roll}{\chi_{\rm roll}}

\newcommand{\D}{\mathrm{d}}

\newcommand{\Cpaper}{\mathbb{A}}
\newcommand{\Dpaper}{\mathbb{B}}

 \newcommand{\Hpaper}{\mathbb{D}}
\newcommand{\Ypaper}{\mathbb{E}}
 
 \newcommand{\AApaper}{{\mathbb
    F}}
 \newcommand{\BBpaper}{{\mathbb G}}
\newcommand{\CCpaper}{{\mathbb H}}
 \newcommand{\DDpaper}{{\mathbb
    I}}
 \newcommand{\EEpaper}{{\mathbb J}}
\newcommand{\YYpaper}{{\mathbb K}}

\definecolor{gnuplot-ls1}{HTML}{9400d3}
\definecolor{gnuplot-ls2}{HTML}{009e73}
\definecolor{gnuplot-dark-red}{HTML}{8b0000}

\title{U-shaped disks in Stokes flow: Chiral
  sedimentation of a non-chiral particle}

\author{Christian Vaquero-Stainer\aff{1},
  Tymoteusz Miara\aff{2},
  Anne Juel\aff{2},
  Draga Pihler-Puzovi\'{c}\aff{2},
    Matthias Heil\aff{3}}

\affiliation{\aff{1}{Department of Physics and  Astronomy, University of Manchester, Manchester M13 9PL,
  U.K. Now at: Nonlinear and Non-equilibrium Physics Unit, OIST
  Graduate University, Onna 904-0495, Japan}\aff{2}{Department of Physics and
    Astronomy, University of Manchester, Manchester M13 9PL, U.K.}
  \aff{3}{Department of Mathematics, University of Manchester, Oxford Road, Manchester M13 9PL, U.K.}}

\usepackage[symbol]{footmisc}

\begin{document}
 
      \maketitle

      \begin{abstract}
 We study the sedimentation of U-shaped circular disks in the Stokes
limit of vanishing inertia. We simulate the flow past
such disks using a finite-element-based solution of the 3D
Stokes equations, accounting for the integrable singularities that
develop along their edges. We show that the purely vertical sedimentation
of such disks in their upright- [upside-down-] U orientation is
unstable to perturbations about their pitching [rolling]
axes. The instability is found to depend only weakly on the size of the
container in which the disks sediment, allowing us to analyse their
behaviour based on the resistance matrix which governs the
evolution of the disk's six rigid-body degrees of freedom in an
unbounded fluid. We show that the governing equations can be reduced
to two ODEs which describe the disk's inclination against the
direction of gravity. A phase-plane analysis, results of which are
in good agreement  with experiments, reveals that the two
instabilities generally cause the disk to sediment along complex
spiral trajectories while it alternates between pitching- and rolling-dominated
motions. The chirality of the trajectories is set by the initial
conditions rather than the (non-chiral) shape of the disk. For certain
initial orientations, the disk retains its inclination and sediments
along a perfectly helical path. 
The observed behaviour is fundamentally different from that
displayed by flat circular disks which sediment without any
reorientation. We therefore study the effect of variations in the disk's
curvature to show how in the limit of vanishing curvature 
the behaviour of a flat disk is recovered.
      \end{abstract}

\section{\label{sect:intro}Introduction}

The sedimentation of particles in a viscous fluid is
a classical problem in fluid mechanics, and plays an important
role in many industrial and natural processes. For example, the fate
of microplastics found on the seabed is a prominent environmental
question, which requires predictions of particle transport, deposition
and resuspension \citep{Claessens13, Turner21}. In many applications the flow induced by 
sedimenting particles is inertially dominated and much of the particles'
complex behaviour is due to the associated nonlinear effects; see
e.g.~\cite{ErnEtAl2012,Heisinger14} for an overview of different
behaviours exhibited by planar disks sedimenting at finite Reynolds
number. However, handling and processing of microscale materials,
such as ultrathin graphene flakes and other colloidal objects
in liquid environments typically involves flows at
small Reynolds number~\citep{Khan2012, Ma2021, Swan21}.
In the same inertialess regime, single particle sedimentation  
has been utilised for understanding the 
size-separation of DNA and other biological
proteins~\citep{Weber13},
and improving clinical blood tests \citep{Peltomaki13}.

In the Stokes limit of vanishing inertia, the sedimenting particle's
shape is the key factor determining its motion, and the behaviour of
many simple objects in unbounded fluids is well understood. For
instance, at zero Reynolds number spheres sediment purely
vertically~\citep{Stokes51}, while straight rods and flat circular
disks sediment with a constant horizontal drift and without changing
their orientation~\citep{Brenner63}. Conversely, chiral particles (such
as a propeller-shaped objects) continually re-orient while
sedimenting, resulting in spiral trajectories \citep{Doi2005,
  Witten2020}. The sense of rotation
of such screw-like objects is determined by their geometry; see
\cite{Gonzalez2004} for a comprehensive
analysis. However, despite the existence of a well-established
theoretical framework, which uses the mobility or
resistance matrix to
link the particle's translational velocity and its rate of rotation to the
forces and torques acting on it~\citep{HappelAndBrenner}, predicting
the qualitative behaviour of an arbitrarily shaped sedimenting
particle in an infinite fluid at zero Reynolds is still an open
question \citep{Huseby24, Collins21}.

The behaviour of individual particles has an important effect on their
collective behaviour in dilute suspensions, i.e. in a regime when
interactions between particles can be neglected. For instance, a
dilute cloud of spheres will sediment without changing its shape while
dilute clouds of straight rods or planar disks will disperse because
each individual particle will move in a different direction
and with a different speed, depending on the particles' random initial    
orientations. To the best of our knowledge, all chiral objects investigated so far 
sediment with nonzero horizontal
velocities, but their spiral trajectories imply that, on average,
they sediment purely vertically \citep{Witten2020, Huseby24}. This
implies that a dilute cloud of such objects sediment with net
zero horizontal dispersion.

If the sedimenting particles are sufficiently flexible, the
deformation induced by the fluid traction can change
their behaviour. For instance, sedimenting elastic fibres tend to
deform into a U-shape and then right themselves until they reach an
upright-U orientation, following which they sediment steadily without
any horizontal drift -- unlike their rigid counterparts.
The sedimentation of such flexible fibres has been studied
extensively (see, e.g., \cite{Spagnolie2013, Marchetti2018}),
but there is far less work on the
effect of fluid-structure interaction on the sedimentation of other
objects. Recent experiments of \cite{MiaraThesis} and numerical
situations of \cite{Yu23}, showed
that sedimenting flexible sheets can also deform into a
U-shape, similar to the shape adopted by sedimenting fibres.

\begin{figure}
\begin{center}
\includegraphics{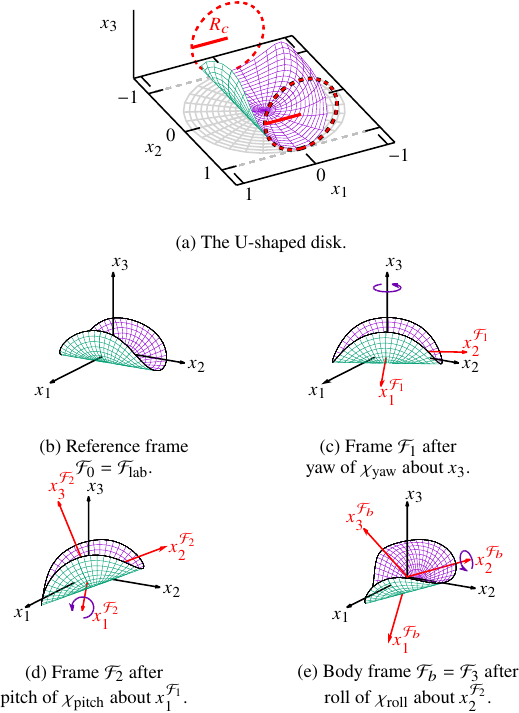}
  \caption{(a) Schematic of the isometric deformation of a flat circular
    disk onto the surface of a cylinder of radius $R_c$. (b-e) Tait-Bryan
    angle convention following the intrinsic
    yaw$\rightarrow$pitch$\rightarrow$roll sequence, showing (b)
    the reference frame $\mathcal F_0$; (c)-(d) the orientation in
    the intermediate frames $\mathcal F_1$ and $\mathcal F_2$;
    (e) the body frame $\mathcal F_b$.
  \label{fig:setup}}
\end{center}
\end{figure}

Motivated by these observations, \cite{Miara2024} performed an
experimental study of the sedimentiation
of such objects, focusing on the behaviour of rigid U-shaped disks which we
created by the isometric deformation of a circular disk onto
the surface of a circular cylinder, as shown
in figure \ref{fig:setup}(a). The resulting U-shaped disk has two planes
of symmetry and, crucially, is non-chiral. We studied the sedimentation of
such disks in a viscous fluid, with the viscosity and density
difference chosen such that the resulting flow had a small Reynolds
number. Our experiments showed that over the experimentally observable
lengthscales (limited by the size of the tank in which the experiments
were performed), such disks never settled into a steady motion but
continued to reorient, with concomitant continuous changes to their
velocity. A careful analysis of the observed trajectories showed that
this behaviour is due to the disks undergoing a periodic sequence of
rolling and pitching motions (see figure \ref{fig:setup} b-e for an illustration
of the terminology and the angles used to describe the disk's
orientation) which resulted in a sedimentation along complicated spiral
trajectories whose chirality is determined by the disk's initial
orientation, rather than being set by the (non-chiral) particle
shape \citep{Miara2024}. 

The aim of the present paper is to analyse this behaviour using a
combination of numerical and analytical approaches, in order to
elucidate the underlying physical mechanisms and the resulting
dynamics. We initially formulate
the problem (based on the solution of the 3D Stokes equations, coupled
to the dynamics of the sedimenting U-shaped disk) in a finite
domain, mimicking the geometry of the finite-sized
tank used in the experiments. We assume that the disk sediments
quasi-steadily so that the body force due to the (negative) buoyancy
exactly balances the fluid traction.
We then use numerical simulations to assess the stability of the disk
when it sediments purely vertically in an upright-U orientation
along the centre of a finite cubic tank. This shows that, in
this orientation, the disk is stable to perturbations about its roll
axis (similar to the behaviour observed for sedimenting U-shaped
fibres which re-orient towards their upright-U orientation). However,
the disk is found to be unstable to perturbations about its pitching
axis. We show that this behaviour is only weakly affected by the walls of the
finite computational domain. This allows us to analyse the disk's
behaviour in an unbounded fluid, using an approximate resistance
matrix obtained from a small number of numerical solutions of the full
3D Stokes equations in our finite computational domain. We reduce
the six coupled ODEs which govern the disk's motion and reorientation
to a system of two coupled ODEs for the evolution of the disk's
roll and pitch angles. We then employ a phase-space analysis to elucidate
the disk's behaviour and thus explain the mechanism responsible for
the spiral trajectories observed in the experiments.

\section{Problem setup}

We consider a thin disk of area $A$, nominal uniform thickness $h$ and
homogeneous density $\rho_{\rm d}$, sedimenting in a fluid of
viscosity $\mu$ and density $\rho_{\rm f} = \rho_{\rm d} -
\Delta \rho$, where $\Delta \rho > 0$, with gravity of strength
$g$ acting in the negative $x_3$-direction. We non-dimensionalise all
lengths on the characteristic length ${\cal L} = (A/\pi)^{1/2}$,
the velocity on ${\cal U} = \Delta\rho g {\cal L}^2/\mu$,
the pressure on the associated viscous scale $\mu
{\cal U}/{\cal L}$ and time on ${\cal L}/{\cal U}$. We restrict
ourselves to the case of small density differences, $\Delta
\rho/\rho_{\rm f} \ll 1$, and assume that the
Reynolds number $Re=\rho_{\rm f} {\cal U}{\cal L}/\mu$ and the Strouhal
number $St = {\cal L}/({\cal U}{\cal T})$ are sufficiently small
that inertia can be neglected. The flow is then governed by the
non-dimensional steady Stokes equations
\be
\label{eqn:stokes}
\nabla p = \nabla^2 {\bf u} - \frac{\rho_{\rm f}}{\Delta \rho} \ {\bf e}_3
\ \ \ \mbox{ and } \ \ \  \nabla \cdot {\bf u} = 0.
\ee
The velocity field is subject to no-slip boundary conditions on the
surface of the sedimenting disk. We treat the disk as two-dimensional
and parameterise the non-dimensional position vector ${\bf r}_{\rm d}$
to its surface, ${\partial\Omega_S}$, by two surface coordinates
$(\xi_1,\xi_2)$ so that ${\bf r}_{\rm d}(\xi_1,\xi_2,t)$.
Denoting the position vector to the disk's centre of mass by
${\bf r}_M(t)$, and the vector which gives the disk's
instantaneous rate of rotation
by ${\pmb \omega}(t)$, the no-slip condition becomes
\be
\label{eqn:no_slip}
 {\bf u}\big({\bf r}_{\rm d}(\xi_1,\xi_2),t\big) =
 \frac{{\rm d}{\bf r}_M(t)}{{\rm d}t} +
 {\pmb \omega}(t) \times \bigg({\bf r}_{\rm d}(\xi_1,\xi_2,t)-{\bf r}_M(t)\bigg).
 \ee
 When studying the sedimentation in a finite-sized container, we apply
no-slip conditions on the container walls and impose zero vertical velocity
and zero tangential stress at the  free surface (which we assume to
remain flat) to match the experimental conditions used in
\cite{Miara2024}.

For modest density ratios, such that $(\rho_{\rm
  d}/\rho_{\rm f}) Re \ St \ll 1$ the disk's inertia can be neglected and
the equations governing the evolution of ${\bf r}_{\rm
  M}(t)$ and ${\pmb \omega}(t)$ are given by
the balance of forces and torques on the disk: the fluid traction has
to balance the net body force, ${\bf F}$ (non-dimensionalised on
$\Delta \rho g {\cal L}^3$), so 
\be
\label{eqn:rigid_body_force_balance_nondim}
\int_{\partial\Omega_S}{\pmb \tau}\cdot\hat{\bf n}\; {\rm d}S = {\bf
  F}.
\ee
The net torque of the fluid traction has to balance any externally
applied torque, ${\bf T}_M$ (non-dimensionalised on
$\Delta \rho g {\cal L}^4$), where we compute both torques about the
disk's centre of mass, so
\be
\label{eqn:rigid_body_torque_balance_nondim}
\int_{\partial\Omega_S}
    \big({\bf r}_{\rm d}- {\bf r}_M \big) \times
    \big({\pmb \tau}\cdot\hat{\bf n}\big)\;{\rm d}S = {\bf T}_M.
\ee
Here we assume a Newtonian constitutive relation, and so the
components
of the stress tensor ${\pmb \tau}$ are given by
$\tau_{ij} = -p \delta_{ij} + ( \partial u_i/\partial x_j + \partial
u_j/\partial x_i)$, the vector $\hat{\bf n}$
is the outer unit normal to the disk, and the surface $\partial \Omega_S$ encompasses 
both sides of the disk. 
For a freely sedimenting disk, the net body force is due to buoyancy,
${\bf F} = -\pi h/{\cal L} \; {\bf e}_3$, and there is no net
torque, ${\bf T}_M = {\bf 0}$.

If the container is sufficiently large so
that the effect of its walls on the sedimentation can be neglected, the
velocities decay to zero and the pressure gradient approaches a purely
hydrostatic distribution,
$\nabla p \to - (\rho_{\rm f}/\Delta \rho) \ {\bf e}_3$,
at large distances from the disk. In that case, the velocity ${\bf U}_M$ of the
disk's centre of mass, and its rate of rotation about its centre of
mass, ${\pmb \omega}$, are
determined by
\be
\begin{pmatrix}
  {\bf F} \\
  {\bf T}_M
\end{pmatrix} =
{\bf R}\cdot 
\begin{pmatrix}
  {\bf U}_M \\
  {\pmb \omega}
\end{pmatrix},
\label{eqn:fsi_governing_eqn_nondim}
\ee
where ${\bf R}$ is a $6\times 6$ symmetric positive-definite matrix known as 
the resistance matrix \citep{HappelAndBrenner}.

We decompose the various vectors in
(\ref{eqn:fsi_governing_eqn_nondim}) into a body-fitted coordinate system,
${\cal F}_b$, aligned with the disk's principal axes,
$({\bf e}_1^{{\cal F}_b},{\bf e}_2^{{\cal F}_b},{\bf e}_3^{{\cal
    F}_b})$, such that, e.g.
${\bf U}_M =
U_{1}^{{\cal F}_b} \, {\bf e}_1^{{\cal F}_b} +
U_{2}^{{\cal F}_b} \, {\bf e}_2^{{\cal F}_b} +
U_{3}^{{\cal F}_b} \, {\bf e}_3^{{\cal F}_b};$
see figure \ref{fig:setup}(e). We employ the Tait-Bryan
angles $\yaw, \pitch$ and $\roll$ which describe an
intrinsic sequence of rotations to define the
orientation of the disk relative to the lab frame.
Since rotations do not commute, it is important to remember
throughout this paper that the rotations have to be
applied consecutively in the order
yaw$\rightarrow$pitch$\rightarrow$roll. This rotates
the disk from the lab frame ${\cal F}_{\rm lab} =
{\cal F}_{0}$ through the intermediate frames 
${\cal F}_{1}$ and ${\cal F}_{2}$ to the body-fitted
frame ${\cal F}_{3} = {\cal F}_b$, as illustrated in
figures \ref{fig:setup}(b-e); see appendix
\ref{sec:angle_stuff} for details.

In the $({\bf e}_1^{{\cal F}_b},{\bf e}_2^{{\cal F}_b},{\bf e}_3^{{\cal
    F}_b})$-coordinate system the resistance matrix is constant
and, in general, contains 21 independent entries, encoded by the three
tensors ${\bf K}={\bf K}^T, {\bf \Omega}_M =
{\bf \Omega}_M^T$ and ${\bf C}_M$ (non-dimensionalised on
$\mu {\cal L}, \mu {\cal L}^2$ and $\mu {\cal L}^3$, respectively). 
We thus have
\be
\bigg[
  F_1^{{\cal F}_b},
  F_2^{{\cal F}_b},
  F_3^{{\cal F}_b},
  T_{1}^{{\cal F}_b},
  T_{2}^{{\cal F}_b},
  T_{3}^{{\cal F}_b}
  \bigg]^T =
\begin{pmatrix}
  {\bf K} & {\bf C}_M^T \\
  {\bf C}_M & {\bf \Omega}_M
\end{pmatrix}
\bigg[
  U_{1}^{{\cal F}_b},
  U_{2}^{{\cal F}_b},
  U_{3}^{{\cal F}_b},
  \omega_{1}^{{\cal F}_b},
  \omega_{2}^{{\cal F}_b},
  \omega_{3}^{{\cal F}_b}
  \bigg]^T.
\label{eqn:fsi_governing_eqn_nondim2}
\ee
To aid the clarity of our subsequent discussions
we will generally refer to the components
$T_{\{1,2,3\}}^{{\cal F}_b}$ of the external torque acting
on the disk about its body-fitted $x^{{\cal F}_b}_{\{1,2,3\}}$-axes
as  $T_{\{{\rm pitch,roll,yaw}\}}^{{\cal F}_b}$.
The translation tensor ${\bf K}$ is
symmetric and characterises the hydrodynamic drag induced by
translational motion along the principal axes; the rotation tensor
${\pmb\Omega_M}$ is symmetric and characterises the hydrodynamic
torque induced by rotational motion about the principal axes at the
centre of mass; and the coupling tensor ${\bf C_M}$, non-symmetric in
general, characterises translation-rotation coupling, i.e. the torque
induced by a translational motion and the drag induced by a rotational
motion. The resistance matrix is symmetric as a consequence of
the symmetry of $\bm K$ and $\bm \Omega_M$, and, in the frame ${\cal
  F}_b$, depends only on the geometry of the body. For the
doubly-symmetric disks considered in this study, only a subset of the
entries in ${\bf R}$ is nonzero; see appendix
\ref{sec:grand_resistance_matrix} for details.

\section{Numerical solution\label{sec:numerical_solution}}
\subsection{\label{sec:augmented_stokes}Solution of the quasi-steady Stokes equations}
\begin{figure}
\begin{center}
\includegraphics{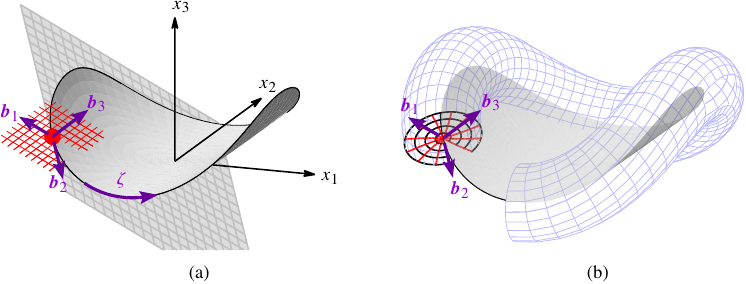}
\caption{\label{augmentation_sketch}(a) Illustration of the tangent
  plane to the disk whose edge is parametrised by the coordinate
  $\zeta$. The three orthogonal unit vectors ${\bf b}_1, {\bf b}_2,
  {\bf b}_3$ are chosen such that
  ${\bf b}_2$ is tangential to the edge of the disk; ${\bf b}_1$ is
  tangential to the disk but normal to its edge; ${\bf b}_3 = {\bf b}_1
  \times {\bf b}_2$ is normal to the disk and its edge. (b) The
  toroidal region within which the finite element
  solution is augmented by suitable singular functions; red lines
  are contours of constant $\theta$, and black circular lines and contours
  of constant $\rho$.}
\end{center}
\end{figure}

Given the disk's position and orientation, and their instantaneous
rates of change, characterised by ${\rm d}{\bf r}_{M}/{\rm d}t$
and $\pmb{\omega}$, we solved the Stokes equations (\ref{eqn:stokes}),
subject to the kinematic boundary condition (\ref{eqn:no_slip}) and the 
no-slip/traction-free-surface conditions on the outer boundaries of the fluid
domain by a finite-element method, using tetrahedral Taylor-Hood
elements on an unstructured body-fitted mesh, generated with {\tt
  Tetgen} \citep{tetgen}. A challenge arises from the presence
of the (integrable) singularities in the pressure and shear stress
along the edge of the disk \citep{Gupta57, Tannish96}.
These contribute significantly to the
hydrodynamic drag and torque acting on the disk. For instance, for a flat disk,
sedimenting in broadside orientation,
30\% of the drag is generated in the outermost 5\% of the disk's radius. The
singularities cannot be resolved using standard finite
elements, even if adaptive mesh refinement is applied. Therefore, we augmented
the standard finite element basis functions by appropriate
singular functions. For a smoothly
deformed circular disk, such as the one shown in
figure \ref{augmentation_sketch}, the flow field in the vicinity of its
curved edge, whose position we describe by the vector
${\bf r}_{\partial D}(\zeta)$, resembles the flow past the straight
edge of a semi-infinite flat plate oriented tangential to the
disk, as shown in figure \ref{augmentation_sketch}(a). We therefore introduce a
body-fitted toroidal coordinate system $(\zeta,\rho,\theta)$ that
is aligned with the edge of the disk and defined relative to the
three orthogonal unit vectors ${\bf b}_1, {\bf b}_2, {\bf b}_3$. Here
${\bf b}_2 = (\D{\bf r}_{\partial D}/\D\zeta)/|\D\bm r_{\partial D}/\D\zeta|$ is
tangential to the edge of the disk; ${\bf b}_1$ is
tangential to the disk but normal to its edge; ${\bf b}_3 = {\bf b}_1
\times {\bf b}_2$ is normal to the disk and its edge. The coordinates
are chosen so that $\rho \ge 0$ and
$0 \le \theta \le 2\pi$  describe the position within a radial slice
whose normal is ${\bf b}_2$, as shown in
figure \ref{augmentation_sketch}(b). In a finite toroidal region
surrounding the edge of the disk we then represented the velocities
and pressure as
   \be
   \big\{ {\bf u}, \ p \big\} =
   \bigg\{{\bf u}^{\rm[FE]}, \ p^{\rm [FE]}\bigg\} + \sum_{i=1}^{3} C_i(\zeta) \ 
   \bigg\{ {\bf u}^{\rm [sing]}_{i}(\rho,\theta),
   \ p^{\rm [sing]}_{i}(\rho,\theta)\bigg\},
   \ee
where the ${\bf u}^{[sing]}_{i}, p^{[sing]}_{i}$ are the velocity and
pressure fields arising from a translation of a (planar straight) edge
with unit velocity in the local ${\bf b}_i$ direction; they are
scaled by the a-priori unknown amplitudes $C_i(\zeta)$. Outside this
region, the velocities and pressures were represented by the
finite-element basis functions alone, with continuity across the
interface between the augmented and non-augmented regions imposed by
Lagrange multipliers. We expanded the amplitudes $C_i(\zeta)$ using
one-dimensional Hermite finite elements and determined their discrete
amplitudes by formulating the entire problem as a PDE-constrained
optimisation problem which ensured that within the augmented region,
the finite element part of the solution $\big\{{\bf u}^{\rm[FE]}, \ p^{\rm [FE]}\big\}$
was as smooth as
possible. This was done by minimising the $L^2$-norm of the
inter-element jump in the traction associated with the finite element
part of the solution across the faces of adjacent tetrahedal
elements in the augmented region. Full details of the method, which
was implemented in {\tt oomph-lib} \citep{HeilHazelOomph2006, oomph-lib_2022},
are provided in
\cite{VaqueroStainerThesis,ChristianNumericsInPreparation}.

Given the velocity and pressure fields, equations
(\ref{eqn:rigid_body_force_balance_nondim}) and
(\ref{eqn:rigid_body_torque_balance_nondim}) then provide
six implicit ODEs for the rate of change of the vector to the
disk's centre of mass, ${\bf r}_M(t)$, and the disk's orientation which
could, in principle, be time-integrated to obtain the motion of the
sedimenting disk. 

\subsection{\label{sect:6x6}The disk's motion in an unbounded fluid}
We will show below that for a sufficiently large container (or
computational domain), boundary effects
are sufficiently weak so that the hydrodynamic drag and torque
acting on the sedimenting disk are close to those experienced by a disk
sedimenting in an unbounded fluid. Equation
(\ref{eqn:fsi_governing_eqn_nondim2}) then allows us to
determine an approximation of the resistance matrix ${\bf R}$
by performing six computations for the flow past that disk in a finite
computational domain: we
positioned the disk so that its centre of mass coincided with the
origin of the lab coordinate system, ${\cal F}_{\rm lab}$, centred
in the middle of the container, and its principal axes coincided with
the coordinate axes, ${\cal F}_{\rm lab} = {\cal F}_{\rm b}$. For
each solve we then set one of the six components on the right-hand side of
(\ref{eqn:fsi_governing_eqn_nondim2}) to one, while setting the others
to zero. Solving the Stokes equations with the corresponding boundary
conditions (\ref{eqn:no_slip}) then determined the velocity and
pressure fields in the fluid, and,
via (\ref{eqn:rigid_body_force_balance_nondim})
and (\ref{eqn:rigid_body_torque_balance_nondim}), the net drag, ${\bf F}$,
and torque, ${\bf T}_M$ acting on the disk. The six components of
these two vectors thus determined an approximation to
one column of the $6 \times 6$ resistance matrix ${\bf R}$. We
refer to appendix \ref{sec:wall_effects} for an assessment of the
errors introduced by this approximation.

Given the entries of the resistance matrix ${\bf R}$, we
describe the motion of the sedimenting disk in an unbounded fluid
in terms of the position vector to its centre of mass, ${\bf r}_M=
\bigg[r_{M,1}^{{\cal F}_{\rm lab}}, r_{M,2}^{{\cal F}_{\rm lab}},r_{M,3}^{{\cal F}_{\rm lab}}\bigg]$, and its
orientation, the latter described by the Tait-Bryan angles
$\yaw, \pitch$ and $\roll$ introduced in
figure \ref{fig:setup}. We regard $\yaw = \roll =\pitch=0$ as the
reference state in which the disk is in an upright-U configuration,
with its principal axes aligned with the lab frame. 
Transforming equation (\ref{eqn:fsi_governing_eqn_nondim2}) from its
body-fitted coordinate system, ${\cal F}_b$, to the lab-frame,
${\cal F}_{\rm lab}$, then yields a system of six autonomous ODEs of the form
\be
\label{eqn:ode_system}
\frac{{\D}{\bf X}}{\D t}  =
{\bf f}\left({\bf X}\right) 
\ee
where
${\bf X} = \bigg[
  r_{M,1}^{{\cal F}_{\rm lab}},
  r_{M,2}^{{\cal F}_{\rm lab}},
  r_{M,3}^{{\cal F}_{\rm lab}},
  \yaw,
  \pitch,
  \roll \bigg] $; see appendix \ref{sec:angle_stuff} for
details. The solution of these ODEs is subject to initial conditions for
all six quantities, ${\bf X}(t=0) = \bigg[
  r_{M,1}^{[0]{\cal F}_{\rm lab}},
  r_{M,2}^{[0]{\cal F}_{\rm lab}},
  r_{M,3}^{[0]{\cal F}_{\rm lab}},
  \yaw^{[0]},
  \pitch^{[0]},
  \roll^{[0]}  \bigg]$ but it is easy to see that the dynamics
are only affected by the initial values of the roll and pitch
angles which quantify the inclination of the disk. This is because in an unbounded fluid of homogeneous density the disk's motion is not
affected by a change in the initial position of its centre of mass: the
resulting trajectory is simply subject to a constant rigid body
displacement. Similarly, a change in the initial yaw angle rotates
the disk about an axis parallel to the direction of gravity and
the entire subsequent motion simply inherits this constant rigid
body rotation; the reorientation dynamics are therefore unaffected.

We show in appendix \ref{sec:angle_stuff} that the evolution of $\pitch$ and
$\roll$ is governed by two coupled, autonomous
ODEs, given by
\be
\label{pitch_and_roll_odes}
\frac{\D\pitch}{\D\tilde{t}} = f_{\rm pitch}(\pitch,\roll) \mbox{\ \ \ and \ \ \ } 
\frac{\D\roll}{\D\tilde{t}} = f_{\rm roll}(\pitch,\roll),
\ee
where 
\be
\label{pitch_ode_rhs}
 f_{\rm pitch}(\pitch,\roll) =
\frac{\Cpaper}{\Ypaper} \sin(\pitch) \cos(\roll)
\ee
and 
\be
\label{roll_ode_rhs}
f_{\rm roll}(\pitch,\roll) = 
\frac{1}{\Ypaper} \frac{\sin(\roll)}{\cos(\pitch)}
\left(
\Dpaper \cos^2(\pitch) - \Hpaper \right).
\ee
Once these ODEs are solved for $\pitch(t)$ and
$\roll(t)$, the evolution of the yaw angle follows from
\be
\label{yaw_ode}
\frac{\D\yaw}{\D\tilde{t}} = f_{\rm yaw}(\pitch,\roll) = 
\frac{\Hpaper}{\Ypaper} \frac{\sin(\roll)\sin(\pitch)}{\cos(\pitch)},
\ee
and the trajectory of the disk's centre of mass can be obtained from
\be
\label{dr_Mdt_ode}
\frac{\D{\bf r}_M}{\D\tilde{t}} =
     {\bf f}_{{\bf r}_M}(\yaw,\pitch,\roll),
     \ee
where
{\small
  $$
 {\bf f}_{{\bf r}_M}(\yaw,\pitch,\roll)
 =
 $$
 \be
 \frac{1}{\YYpaper}
 \left(
\begin{array}{l}
 -\cos(\pitch) \big(
\AApaper \sin(\yaw) \sin(\pitch) \cos^2(\roll) +
\BBpaper \cos(\yaw) \sin(\roll) \cos(\roll) +
\CCpaper \sin(\yaw) \sin(\pitch)\big)
\\
\ \ \ \cos(\pitch) \big(
\AApaper   \cos(\yaw)  \sin(\pitch) \cos^2(\roll)
- \BBpaper \sin(\yaw)  \sin(\roll)  \cos(\roll)  
+ \CCpaper \cos(\yaw) \sin(\pitch) \big)
\\
 \EEpaper - ( \AApaper \cos^2(\roll) + \DDpaper) \cos^2(\pitch)
\end{array}
\right).
\ee
}
Here, the constants
$\mathbb{A},...,\mathbb{K}$ are functions of the entries in
the resistance matrix (see appendix
\ref{sec:grand_resistance_matrix}), and time was rescaled as $\tilde{t} = (\pi h/{\cal L}) \ t$.

\section{Results}
\label{sec:results}
We initially present results obtained for a circular disk bent
isometrically into a cylindrical U-shape with a constant radius of
curvature $R_c=0.5$, sedimenting in a cubic container with dimensions
${\tt L} \times {\tt L} \times {\tt L}$ where ${\tt L} = 160$. 

\subsection{The stability of purely vertical sedimentation}
\begin{figure}
\begin{center}
\includegraphics{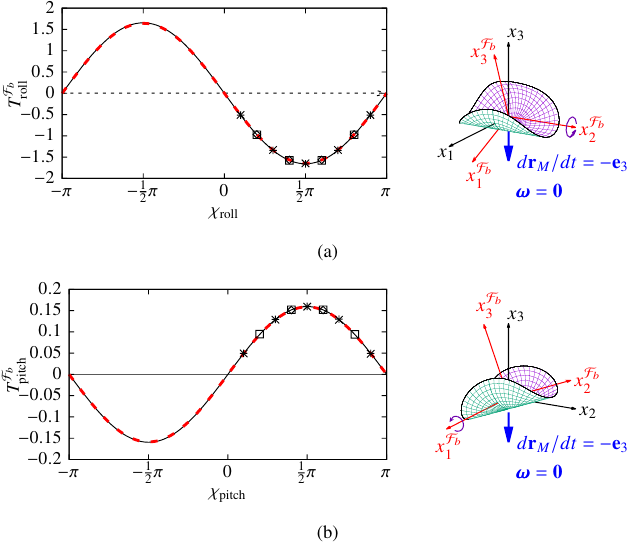}  
\caption{\label{torques}Torques acting on the U-shaped disk 
    when translating with unit speed in the negative $x_3$-direction.
    (a) Torque $T^{{\cal F}_b}_{\rm roll}$ about the body-fitted roll axis
    $x_2^{{\cal F}_b}$,
    as a function of the roll angle $\roll$ for $\pitch
    = 0.$ (b) Torque $T^{{\cal F}_b}_{\rm pitch}$ about 
    the body-fitted pitch axis, $x_1^{{\cal F}_b}$,
    as a function of the pitching angle $\pitch$ for $\roll
    = 0$. Upright crosses and squares show the results for
    of $\chi_{\rm yaw} = 0$ and $\pi/8$, respectively, in the standard
    computational domain ({\tt L} = 160). Diagonal crosses and
    diamonds show the same data in a domain of twice the size, ({\tt
      L} = 320). The solid and dashed lines show the predictions
    obtained from the resistance matrix, computed when the
    disk in its reference orientation ($\roll=\pitch=\yaw=0$) in a domain of
    size  {\tt L} = 160 and 320, respectively. Disk radius of curvature
    $R_c=0.5$.
}
\end{center}  
\end{figure}

If the disk is released at the centre of the tank while in
its reference (upright U)
or inverted (upside-down U) orientation, with the two symmetry planes
aligned with the vertical container walls, we expect it to sediment purely
vertically and without undergoing any reorientation.
To assess the stability of this motion,
figure \ref{torques} shows the instantaneous hydrodynamic torques
acting on the disk when its centre of mass is positioned at ${\bf
  r}_M(t=0)={\bf 0}$, while it performs a pure vertical translation, with unit
downward velocity, ${\rm d}{\bf r}_M/{\rm d}t=-{\bf e}_3, \ \
\pmb{\omega}= {\bf 0}$.

The `+' symbols in figure \ref{torques}(a) show the 
hydrodynamic torque, $T_{\rm roll}^{{\cal F}_{b}}$,
acting on the disk around its roll axis, $x_2^{{\cal F}_{b}}$, as
the roll angle $\roll$ is varied while keeping the two other
angles fixed at their reference values, $\yaw = \pitch = 0$.
As expected, the torque is zero in the
reference and upside-down orientations,
$\roll=0$ and $\roll=\pi$,
respectively. For all other orientations the torque acts to return the
disk to its reference orientation, and we have 
${\rm d}T_{\rm roll}^{{\cal F}_{b}}/{\rm d}\roll|_{\roll=0} < 0$,
indicating that this orientation
is stable. Conversely, ${\rm d}T_{\rm roll}^{{\cal F}_{b}}/{\rm
  d}\roll|_{\roll=\pi} > 0$, implying that the  upside-down
orientation is unstable.
We note that this behaviour matches that observed 
for U-shaped fibres which typically reorient until they
sediment in an upright-U configuration;
see, e.g., \cite{Spagnolie2013,Marchetti2018}.

The `+' symbols in figure \ref{torques}(b) show the
corresponding data when the disk's orientation is changed
by a rotation about its pitching axis, $x_1^{{\cal F}_{b}}$.
The reference and upside-down
orientations are now characterised by $\pitch=0$ and
$\pitch=\pi$, respectively (note that the upside-down orientation
can be reached by either rolling or pitching). Both orientations
are still equilibria, but, surprisingly, the
reference orientation can be seen to be unstable to pitching
motions in the sense that the hydrodynamic torque acting
on a slightly pitched disk tends to increase $\pitch$
further, ${\rm d}T_{\rm pitch}^{{\cal F}_{b}}/{\rm
  d}\pitch|_{\pitch=0}>0$. Conversely, the upside-down
orientation is stable with respect to such perturbations.  

The other symbols in figure \ref{torques} show the
same data, computed in a container of twice the size 
and for the case when the disk is rotated about the
vertical axis by $\yaw = \pi/8$. The fact that the results
agree extremely well, indicates that the container walls
have very little effect on these torques,
suggesting that when the disk is at the centre of our large but
finite container its response to changes in orientation is
approximately the same as that of a disk moving
in an unbounded fluid. To assess this conjecture, the solid and dashed
lines in figure \ref{torques} show the predictions for the hydrodynamic
torques, $T_{\rm pitch}^{{\cal F}_b}$ and $T_{\rm roll}^{{\cal F}_b}$,
obtained from equation (\ref{eqn:fsi_governing_eqn_nondim2}), using a resistance matrix whose entries
were computed when the disk is in its reference orientation. The
predictions for $T_{\rm roll}^{{\cal F}_{b}}(\roll)$ and
$T_{\rm pitch}^{{\cal F}_{b}}(\pitch)$ obtained from the
solution of the Stokes equations and from the resistance matrix
therefore agree by construction for $\roll = \pitch=0$, but they
can be seen to agree remarkably well over the entire range of possible orientations.

This implies that the reorientation of the sedimenting disk in an
unbounded fluid is well described by the formalism of
\S \ref{sect:6x6}, using a resistance matrix computed in a large-but-finite computational domain.
We adopt this approach in the rest of this paper.
The integration of the six ODEs
(\ref{eqn:ode_system}) then allows us to explore the disk's behaviour
at a fraction of the computational cost that would be required for
the coupled solution of equations (\ref{eqn:stokes}), (\ref{eqn:no_slip})
(\ref{eqn:rigid_body_force_balance_nondim}) and
(\ref{eqn:rigid_body_torque_balance_nondim}), embedded in a
timestepping procedure.

\subsection{The reorientation of the sedimenting disk in an unbounded
  fluid}

\def\disk_inset_scale{0.5}
\begin{figure}
\begin{center}
  \includegraphics{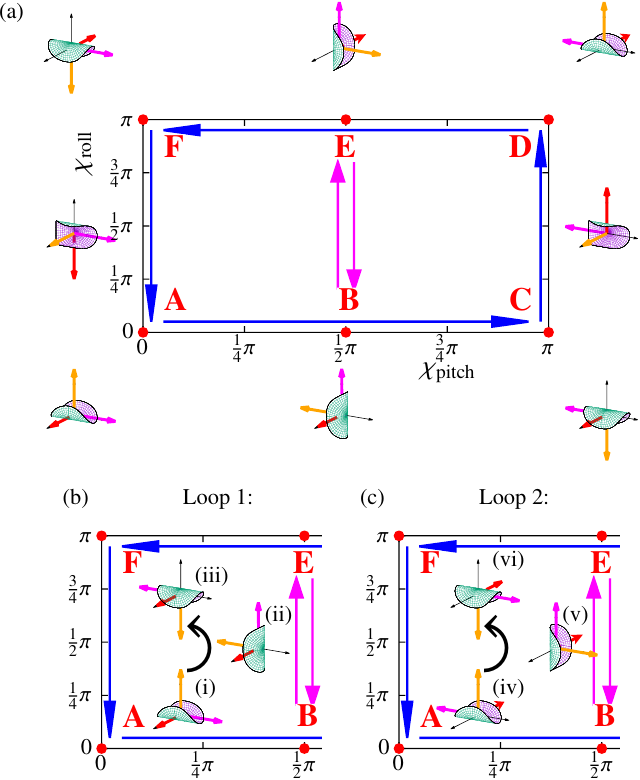}
  \caption{\label{phase_diagram_schematic} (a) Schematic of the
    orientational phase plane spanned by $\pitch$ and $\roll$. The
    insets show representative instantaneous orientations of the
    sedimenting disk which correspond to highlighted points A-F in the
    phase plane. (b,c) Left half of the orientational phase plane
    shown in (a), illustrating the reorientation of a sedimenting disk that starts
    with an initial inclination with small positive values of $\roll$ and
    $\pitch$. Two loops through the $(\pitch,\roll)$-phase plane are required to
    return the disk to its initial orientation.} 
\end{center}
\end{figure}

The study of the disk's sedimentation in an unbounded fluid, based on
its resistance matrix, is facilitated by the fact that,
as shown in \S \ref{sect:6x6}, the disk's dynamics are completely determined
by the evolution of its roll and pitch angles $\pitch$ and
$\roll$, with the remaining quantities, ${\bf r}_M$ and
$\yaw$, being enslaved. This allows us to characterise the
disk's behaviour in the two-dimensional phase plane
shown in figure \ref{phase_diagram_schematic}(a) where
we restrict $\pitch$ and $\roll$
to the range $[0,\pi]$, with the results in the
other three quadrants following from symmetry. The
blue arrows and the insets outside the coordinate axes
in figure \ref{phase_diagram_schematic}(a) illustrate
the disk's reorientation by pure rolling and
pitching. Starting from point A (the reference orientation,
$\yaw = \pitch = \roll = 0$)
a positive perturbation to $\pitch$ initiates a re-orientation by pure
pitching, ${\rm d}\pitch/{\rm d}t>0$, until the disk reaches its
upside-down orientation (point C), as expected from equation
(\ref{pitch_ode_rhs}) and the discussion of
figure \ref{torques}(b). Equations (\ref{roll_ode_rhs})
and (\ref{yaw_ode}) show that the roll and yaw angles remain zero
during this process.

Furthermore, equation (\ref{dr_Mdt_ode}) shows that throughout this motion,
the velocity of the disk's centre of mass only has a component in the lab $x_2$- and
$x_3$-directions. A complete re-orientation, following the path
from point A to C in the phase plane leads to a net zero horizontal
displacement. This is because for $\yaw = \roll = 0$, we have
${\rm d}r_{M,1}/{\rm d}t=0$ and ${\rm d}r_{M,2}/{\rm d}t \sim
\sin(2\pitch)$, therefore the displacement in the positive
$x_2$-direction while the disk pitches through the range
$0< \pitch < \pi/2$ is exactly compensated by a
corresponding negative displacement while
$\pi/2 < \pitch < \pi$. 

If we perturb the disk from its upside-down orientation at
point C ($\yaw = \roll =0; \pitch = \pi$)
by a positive change to $\roll$, the disk re-orients by
pure rolling, ${\rm d}\roll/{\rm d}t>0$, until it approaches the
upright U orientation at point D, now maintaining constant values of the
pitch and yaw angles; see equations (\ref{pitch_ode_rhs})
and (\ref{yaw_ode}). The disk's centre of mass now translates only
in the lab frame $x_1$- and $x_3$-directions and again undergoes a net zero
horizontal displacement as the disk follows the path
from point C to D in the phase plane.

We note that at points A and D, the (symmetric)
disk occupies the same physical space, but its material lines are
rotated by $180^\circ$ about its body-fitted yaw axis; a second
sequence of pitching and rolling motions (from D to F and then back to A)
completely returns it to its reference state. 

Including the three omitted quadrants of the full phase plane
in which both angles vary between $-\pi$ and $\pi$
shows that points A, C, D, F are saddle points, with the stable
directions at saddles A and D corresponding to the disk's
tendency to right itself via rolling, and the unstable directions
corresponding to the tendency to pitch towards an upside-down
orientation. As with any dynamical system, the transition between
two connected saddles would take infinitely long.

We note that, unless $\roll
= 0, \pi$, equations (\ref{roll_ode_rhs}) and
(\ref{yaw_ode}) imply that both ${\rm d}\roll/{\rm d}t$
and ${\rm d}\yaw/{\rm d}t$
blow up as $\pitch \to \pi/2$. This is due to a coordinate singularity
arising from the parametrisation of the disk's orientation in terms of the
Tait-Bryan angles: when $\pitch=\pi/2$ a positive rotation about the yaw axis can be exactly
compensated by an equivalent negative rotation about the roll axis
(recall that the rotations are applied consecutively, in the order
yaw-pitch-roll). However, equations (\ref{roll_ode_rhs}) and
(\ref{yaw_ode}) show that
\be
\label{minus_one}
\lim_{\pitch \to \pi/2} \frac{({\rm d}\roll/{\rm d}t)}{({\rm d}\yaw/{\rm d}t)}  = -1, 
\ee
thus the transition along the line from B to E in the phase
plane happens infinitely quickly but without changing the
orientation of the disk.

This suggests that a path that starts near point A,
but with small positive initial values for $\pitch$ and
$\roll$ will follow an anticlockwise path through the phase
plane. This is illustrated in figures
\ref{phase_diagram_schematic}(b,c).
Starting from an orientation shown by inset (i), the disk performs a pitching-dominated
re-orientation (${\rm d}\pitch/{\rm d}t > 0$ while
$0< \roll \ll 1$), $\pitch$ approaches $\pi/2$,
where the path will undergo a rapid transition
close to the line connecting points B and E (inset (ii))
with only small, continuous changes to the disk's orientation
(${\rm d}\roll/{\rm d}t \approx
-{\rm d}\yaw/{\rm d}t \gg 1$, while $0 < (\pi/2-\roll) \ll 1$).
Once $\roll$ approaches $\pi$, the disk
continues its pitching motion, just below the line from E to F
(${\rm d}\pitch/{\rm d}t < 0$ while $0 < (\pi - \roll) \ll 1$)
until it reaches an approximately upside-down orientation close to
point F (inset (iii)). 
(Note that, following the change in yaw and roll angles during the
transition from B to E, a decrease in $\pitch$ continues
to rotate the disk about its body-frame pitching axis $x_1^{{\cal
    F}_b}$; again a consequence of the order in which the rotations
are applied). Once the disk reaches the
upside-down U orientation near F, it is unstable to rolling, resulting in a
rolling-dominated motion (${\rm d}\roll/{\rm d}t < 0$ while
$0< \pitch \ll 1$), returning the disk to an approximately
upright configuration (inset (iv)). Note that this
rolling-dominated reorientation does not return the disk to the
initial orientation shown in inset (i), but to an orientation equivalent to that
at point D -- shown as inset (iv), where all material lines are
rotated by approximately $180^\circ$ about the body-fitted yaw axis. A
second pitching-rolling sequence
((iv)$\rightarrow$(v)$\rightarrow$(vi)$\rightarrow$(i))
is required to return the disk close to its original orientation.

\begin{figure}
\begin{center}
  \includegraphics{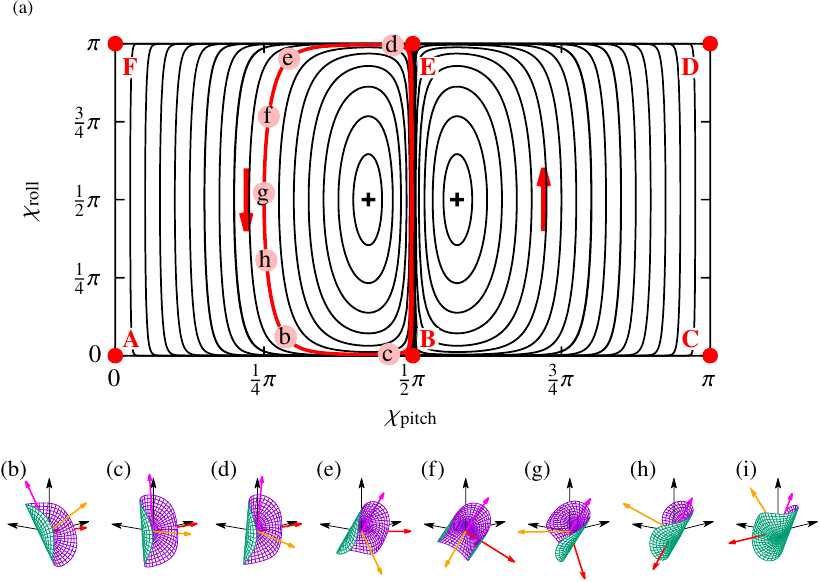}
  \caption{\label{trajectory_with_disk_snapshots} (a) Computed paths in $(\pitch,\roll)$-phase
    plane. (b)-(i) show the disk snapshots while it follows the accentuated path in (a).}
\end{center}
\end{figure}

Figure \ref{trajectory_with_disk_snapshots} shows the actual paths
through the $(\pitch,\roll)$-phase plane
obtained from the numerical integration of equations
(\ref{pitch_and_roll_odes}) for various initial conditions. The plot
confirms the expected behaviour and shows that the paths through the
phase plane form closed anticlockwise loops, implying a periodic reorientation of the
disk during its sedimention. This is illustrated in
figure \ref{trajectory_with_disk_snapshots}(b-i)
which show the disk's orientation in the lab frame while its
inclination ($\pitch(t),\roll(t)$) traces out
the red loop in the phase plane: starting at point (b), the disk
performs a rolling-dominated motion towards point (c) where
the body-fitted roll axis is nearly aligned with the
direction of gravity. As discussed above, there is little change in the
disk's actual orientation as it moves rapidly from points (c) to (d), beyond which it
continues its pitching-dominated motion towards an upside-down orientation.
At point (e), the disk has become sufficiently inverted that
it becomes subject to a strong rolling torque that acts to return it to an
upright position via points (f-i). In the phase plane, point (i) is close to the
starting point (b), indicating the completion of one complete
pitch-roll cycle. The plots of the disk's orientation in the
lab frame in figure \ref{trajectory_with_disk_snapshots}(b,i) confirm
that at point (i) the disk does indeed have the same inclination as at
point (b), but also that it has rotated about the direction of gravity.
The net rotation of the disk is determined by the integration of equation
(\ref{yaw_ode}) for the yaw angle $\yaw(t)$, given the
solution $\pitch(t)$ and $\roll(t)$ from equations
(\ref{pitch_and_roll_odes}). In general, the net change in the yaw
angle during one pitching-rolling cycle is not
equal to a rational multiple of $\pi$, therefore the disk generally performs
a quasi-periodic motion with incomensurate timescales for the
rotation about the direction of gravity and the periodic change in its
inclination. 

Figure \ref{trajectory_with_disk_snapshots} also shows that, as
suggested by the conceptual plot in figure \ref{phase_diagram_schematic},
paths through the phase plane cannot cross the vertical line
$\pitch=\pi/2$ if the disk is released with a non-trivial
initial roll angle, $0 < \roll^{[0]} < \pi$.
Paths to the left of the line BE, starting from $\pitch^{[0]}<\pi/2$, therefore
perform anticlockwise loops through the phase plane; in the lab frame
the disk rotates in a positive sense about the direction of gravity,
${\rm d}\yaw/{\rm d}t < 0$. The paths obtained when starting
from initial conditions to the right of the line BE, i.e. for
$\pitch^{[0]} > \pi/2$ are also anti-clockwise loops,
consistent with the conceptual plot in figure \ref{phase_diagram_schematic}.
This is because the sense of rotation is imposed by the continuity of
the motion for pure pitching motions (along ABC and DEF,
respectively); the counterintuitive clash between the directions in
which the paths just to the right and left of the line BE are
traversed is due to the coordinate
singularity arising from our representation of the disk's incination
in terms of the Tait-Bryan angles. However, when released with an
initial inclination to the right of the line BE, $\pitch^{[0]} > \pi/2$,
the disk rotates in the opposite direction about the direction
of gravity, i.e. ${\rm d}\yaw/{\rm d}t > 0$.

Since the paths through the phase plane form closed
loops they enclose a centre (marked with a cross) which represents a
fixed point of the ODEs (\ref{pitch_and_roll_odes}).
The coordinates $\left(\pitch^{\rm [centre]},\roll^{\rm
  [centre]}\right)$ of that fixed point can therefore be obtained by solving
the  algebraic equations
$f_{\rm pitch}\left(\pitch^{\rm [centre]},\roll^{\rm [centre]}\right) =0$
and
$f_{\rm roll}\left(\pitch^{\rm [centre]},\roll^{\rm
  [centre]}\right) =0$
which yield
\be
\label{fixed_points}
\pitch^{\rm [centre]} =
\arccos \left( \left(\frac{\mathbb{D}}{\mathbb{B}} \right)^{1/2} \right)
\mbox{ \ \ \ and \ \ \ }
\roll^{\rm [centre]} = \pi/2.
\ee

\subsection*{Comparison with experiments}
\begin{figure}
\begin{center}
  \includegraphics{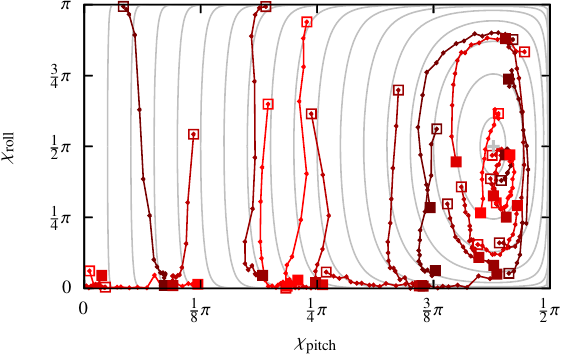}
  \caption{\label{fig:experimental_validation}
    Comparison between the paths in the $(\pitch,\roll)$-phase
    plane obtained numerically (grey lines) and in the experiments of
    \cite{Miara2024} (symbols connected by coloured lines).}
\end{center}
\end{figure}

To assess how well our predictions agree with the experimental
observations of \cite{MiaraThesis, Miara2024},
figure~\ref{fig:experimental_validation} shows
a direct comparison of the paths in the ($\pitch,\roll$)-phase
plane. The symbols connected by coloured lines
represent data from experiments performed with a U-shaped polyamide nylon
disk of density $\rho_s=1130$~kg\:m$^{-3}$, thickness $b=236.7~\mu$m,
area $A=\pi R^2=452$~mm$^2$ and a non-dimensional
radius of curvature of $R_c = 0.525$. The disk was placed in a
cuboidal tank of internal dimensions $900 \times 400 \times 400$~mm$^{3}$ filled
to a height of 750~mm with silicone oil of density
$\rho_f=972.7$~kg\:m$^{-3}$ and dynamic viscosity $\mu =
1.02$~Pa\:s. For each experiment the disk was released with a different
initial orientation, and the subsequent evolution of the
Tait-Bryan angles was then monitored while the disk sedimented over a
vertical distance of approximately $30R$. For the data shown
in figure~\ref{fig:experimental_validation}
the disk's centre of mass was at least $10R$ from the free
surface and the container walls.  Angles outside the range $[0,\pi]$
were mapped into that range, exploiting the disk's symmetry. We
refer to \cite{MiaraThesis, Miara2024} for full details of the experiments. 

Given the finite size of the tank,
each experiment only provides a relatively short path in the phase
plane. However, collectively the paths show good qualitative
agreement with the corresponding theoretical predictions, shown
by the grey lines: the direction of the paths (whose start and end
points are indicated by hollow and filled square symbols,
respectively) is consistent with anticlockwise orbits in the
phase plane. For disks released with modest initial values of $\chi_{\rm pitch}$
and close to an upside-down U orientation ($\chi_{\rm roll}$ close to
$\pi$), the disks reorient rapidly by rolling ($\chi_{\rm roll}
\to 0$) towards an upright U orientation. This is then
followed by a much slower pitching motion, $\chi_{\rm pitch} \to \pi/2$,
while $\chi_{\rm roll}$ remains small. Particles released at larger
initial values of $\chi_{\rm pitch}$ loop around a centre that is
located close to the one predicted by the theory. Small deviations from the
theoretical predictions are visible: e.g. some paths have
${\rm d}\chi_{\rm pitch}/{\rm d}t <  0 [> 0]$ when $\chi_{\rm roll}
<\pi/2  [>\pi/2]$; some paths cross; the centre of the closed orbits is 
slightly below $\chi_{\rm roll} = \pi/2$. However, given the unavoidable
presence of experimental uncertainties, the presence of wall effects (note that the
tank used in the experiments was smaller than the computational
domain in which we computed the resistance matrix that forms
the basis of the theoretical predictions), and imperfections in the
disk shape, the overall agreement is good.

\subsection{\label{sect:labframe}Trajectories in the lab frame of
  reference}

\begin{figure}
 \begin{center}
  \includegraphics{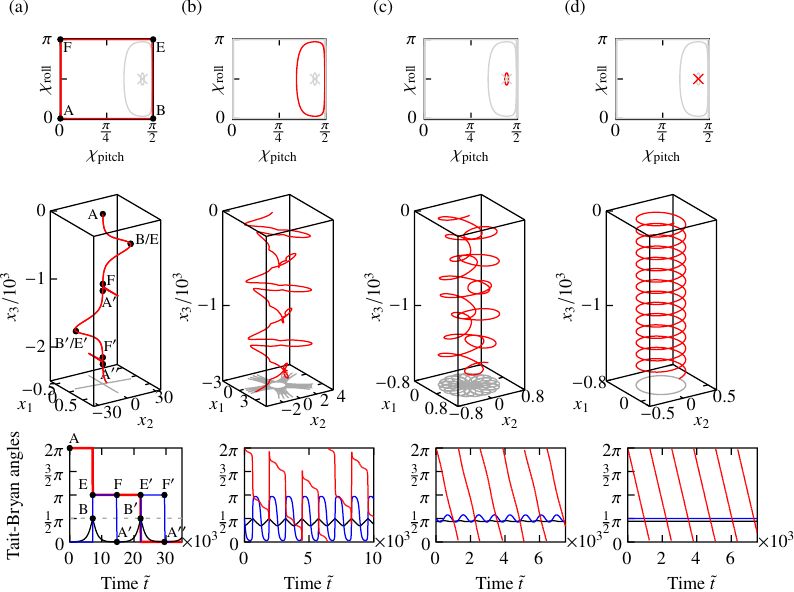}
\caption{The paths in the phase plane (first row) for a disk released at the origin, ${\bf
  r}_{M}^{[0]}={\bf 0}$, with $\yaw^{[0]}=0$, $\roll^{[0]} = \pi/2$
  and (a) $\pitch^{[0]} = 10^{-4}\pi$; (b) $\pitch^{[0]} = \pitch^{[\rm centre]}-0.1\pi$;
 (c) $\pitch^{[0]} = \pitch^{[\rm centre]}-0.01\pi$; (d) $\pitch^{[0]}
  = \pitch^{[\rm centre]}$.
  The corresponding  trajectories of the centre of mass in the lab
  frame of reference
  (middle row) and the time-evolution of the Tait-Bryan angles $\pitch$ (black),
  $\roll$ (blue) and $\yaw$ (red) (bottom row),
  respectively.  \label{com_trajectories}}
 \end{center}
\end{figure}
As discussed above, the dynamics of the disk is completely
encoded by the evolution of its roll and pitch angles, with the remaining
four degrees of freedom (the yaw angle and the position of the
disk's centre of mass) being enslaved to these quantities.

We illustrate the overall motion of the disk in the lab frame in
figure \ref{com_trajectories} where we plot
selected paths in the $(\pitch,\roll)$-phase
space (top), the corresponding  trajectories of the disk's centre of
mass, ${\bf r}_M(t)$, (middle), and the time-traces of the disk's
three Tait-Bryan angles (bottom). In each
case, the disk is released at the origin, ${\bf r}_{M}^{[0]}={\bf 0}$,
with zero yaw angle, $\yaw^{[0]}=0$ (or, equivalently,
$\yaw^{[0]}=2\pi$) and $\roll^{[0]} = \pi/2$. The plots show
the trajectories resulting from four diffent initial values
for the pitch angle. In figure \ref{com_trajectories}(a) we start
with a small value, $\pitch^{[0]}=10^{-4}\pi$, which results in a path close to the outermost boundary
of the accessible phase plane,
and a motion that alternates distinctly between pitching
and rolling, with very little change to the yaw angle. (Note that
${\rm d}\yaw/{\rm d}t|_{t=0} < 0$, therefore the yaw angle
(which we normalised to be in the range $[0,2\pi]$) starts at $\yaw^{[0]}=2\pi$;
similarly $\yaw$ jumps by $2\pi$ whenever $\yaw$ is about to become negative).
Key points on the disk's trajectory are identified by the same labels as in
figures \ref{phase_diagram_schematic} and \ref{trajectory_with_disk_snapshots}:
from A to B, the disk reorients predominantly by pitching; as discussed above,
the rapid, approximately equal and opposite change to the roll and yaw angles when
$\pitch$ is close to $\pi/2$ (from B to E) does not lead to a
significant re-orientation of the disk, and the pitching motion then
continues from E to F. The disk then rights itself by
rolling, via the path from F to A' in the phase plane. The disk
now has retained its original inclination but is rotated by
approximately 180$^\circ$ about the direction of gravity;
a second loop around the phase plane, from A' to A'',
then returns the disk close to its original orientation.

During the pitching dominated phases, the
horizontal drift of the disk's centre of mass is predominantly in the body-frame
$x_2$-direction, while it drifts predominantly in the $x_1$-direction
when reorienting by rolling. Given that at the end of the first loop
around the phase plane (from A to A'), the disk has rotated 
by approximately 180$^\circ$ about the vertical axis,
the direction of the horizontal drift during the second loop (from A'
to A'') occurs in the opposite direction. As a result, the projection
of the trajectory onto the lab's $(x_1,x_2)$-plane (shown with grey
lines in the middle row of images in figure \ref{com_trajectories})
is approximately cross-shaped.

The reorientation by pitching can be seen to occur much more slowly
than by rolling, consistent with the relative magnitudes of the
rolling- and pitching-induced torques shown in figure \ref{torques}, so
the disk spends the majority of its time performing a slow
pitching-dominated motion. During this
phase, most paths in the phase plane (apart from those in the vicinity of the
centre), converge towards a narrow region near $\roll \approx 0,\pi$; see
figure \ref{trajectory_with_disk_snapshots}.
This implies that during this phase, the disk is highly sensitive to perturbations in
the roll angle. Small perturbations introduced by
the numerical integration of the governing equations can therefore
result in large deviations in the paths on later parts of the orbit.
Such numerical perturbations were controlled by performing the
time integration with sufficiently small timesteps; the fact that
the paths close confirms the accuracy of the computations. However,
in lab experiments physical perturbations are
unavoidable, and as a result it is unlikely that such exactly closed
orbits will be observable experimentally. This is consistent
with the experimental observations by \cite{Miara2024}.

In figure \ref{com_trajectories}(b) we start
with a larger initial pitching angle,
$\pitch^{[0]}=\pitch^{[\rm centre]}-0.1\pi$ for which
the trajectory is closer to the interior of the phase plane. Pitching
and rolling therefore occur continuously and the changes in $\pitch$ and
$\roll$ induce a concomitant change in the yaw
angle. Again $\roll$ and $\yaw$ undergo rapid
opposing changes without any significant re-orientation of the disk
whenever $\pitch$ approaches $\pi/2$.

In figure \ref{com_trajectories}(c) we start with a pitching angle
close to the centre, $\pitch^{[0]}=\pitch^{[\rm centre]}-0.01\pi$
where $\pitch^{\rm [centre]} = 0.44\pi$. In the phase plane, the
trajectory therefore remains close to the centre, with the
reorientation dominated by small, periodic oscillations in roll angle,
accompanied by a near constant rate of change in $\yaw$; see
the time-trace of the Tait-Bryan angles at the bottom of the figure.
The changes in roll and pitch angles in figure \ref{com_trajectories}(b,c)
lead to a continual change in the
direction of the horizontal drift, resulting in a spirograph-like
patterns in the projection of the trajectories onto the lab
$(x_1,x_2)$-plane.

Finally, in figure \ref{com_trajectories}(d) the disk is released with 
$\big( \pitch^{[0]},\roll^{[0]}\big) =
\left(\pitch^{\rm [centre]},\roll^{\rm [centre]}\right)$. Pitch and roll angles
therefore remain constant while the yaw angle decreases at a constant
rate, according to ${\rm d} \yaw/{\rm d}t = f_{\rm yaw}\left(\pitch^{\rm [centre]},\roll^{\rm
  [centre]}\right)$. The ODE (\ref{dr_Mdt_ode}) for ${\rm d}{\bf
  r}_M/{\rm d}t$ then implies that the velocity of the disk's centre
of mass varies periodically with the period of the
yawing motion, resulting in a spiral trajectory in the lab frame and
a circular projection onto the lab $(x_1,x_2)$-plane.

\subsection{\label{sec:radius_effects}Dependence on the disk's
  radius of curvature}
So far we have focused on the motion of a U-shaped disk
with a given radius of curvature. We  found the
behaviour of that disk to be fundamentally different from its flat
equivalent: as mentioned in \S \ref{sect:intro}, flat circular disks
that sediment in an unbounded fluid at zero Reynolds number neither
rotate nor change their inclination, and continue to drift in a
fixed horizontal direction while sedimenting. The question
therefore arises how these different behaviours can be reconciled
in the limit as $R_c \to \infty$, i.e. when the U-shaped disk
turns into a flat one.

\begin{figure}
  \begin{center}
  \includegraphics{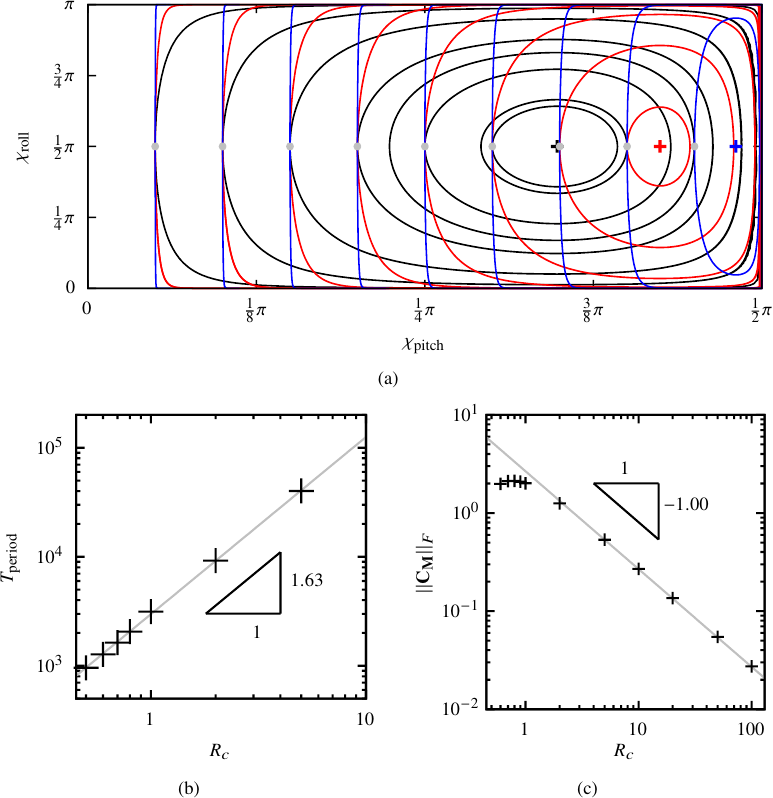}
  \caption{\label{radius_variations}(a) Plot of the paths in the
    $(\roll,\pitch)$-phase plane for disks of different radii of
    curvature (black: $R_c = 0.35$; red: $R_c = 0.5$; blue: $R_c = 2$).
    (b,c) Plots of (b) the time $T_{\rm period}$ for the periodic reorientation of the
    sedimenting disk, and (c) the Frobenius norm of the
    coupling tensor, both as a function
    of the disk's radius of curvature, $R_c$.}
\end{center} 
\end{figure}

Figure \ref{radius_variations}(a) shows that, qualitatively, the
phase plane describing the disk's inclination
remains the same when we change the disk's radius of
curvature: the inclination still undergoes periodic changes,
with the paths in the $(\pitch ,\roll)$-phase plane
following closed orbits around their respective centres. However,
the rate at which the disk reorients decreases with an increase in
$R_c$. This is illustrated in  figure \ref{radius_variations}(b) where we plot,
\be
T_{\rm period} = 2\pi \ \left( \left.
\frac{{\rm d}\yaw}{{\rm d}t} \right|_{\left(\pitch^{\rm [centre]},\roll^{\rm [centre]}\right)} \right)^{-1},
\ee
which is the time it takes for the disk to rotate by 360$^\circ$ about
the direction of gravity when
sedimenting along its spiral path at a fixed inclination. The solid
line is a power-law curve fit to the computational data and suggests
that $T_{\rm period} \sim R_c^{1.63}$ as $R_c \to \infty$.
Thus, nearly flat disks (with a very large radius of curvature) still
move around their periodic orbits in the phase plane, but they
do so increasingly slowly, so that their inclination remains
approximately (and, in the actual limit $R_c \to \infty$, exactly)
constant.

This behaviour is reflected in the fact that the entries in the
off-diagonal coupling block, ${\bf C}_M$, in the resistance
matrix go to zero as the radius of curvature increases. This is
illustrated in
figure \ref{radius_variations}(c) by the plot of the Frobenius norm
$||{\bf C}_M||_F = (\sum_i\sum_jC_{M,ij}^2)^\frac{1}{2}$ as a function
of $R_c$ which can also be seen to follow an
approximate power-law behaviour, with  $||{\bf C}_M||_F \sim
R_c^{-1}$, so that when $R_c \to \infty$ translational motions
no longer induce rotations (and vice versa).

One possible interpretation of these results is that in the
limit $R_c \to \infty$, any initial inclination
$\left(\pitch^{[0]},\roll^{[0]}\right)$ becomes a fixed
point in the sense that the motion along the paths in the
$(\pitch,\roll)$-phase plane becomes infinitely slow.
However, it is also of interest to see
what happens to the centres (which are always fixed points)
as the disk's radius of curvature is
increased. Equation (\ref{fixed_points}) showed that $\roll^{\rm
  [centre]} = \pi/2$, irrespective of the disk's radius of
curvature, while $\pitch^{{\rm [centre]}}$ depends on the
ratio of the coefficients $\mathbb{B}$ and $\mathbb{D}$, both of which
are given in terms of the coefficients of the resistance matrix
in appendix \ref{sec:grand_resistance_matrix}. As $R_c \to \infty$, the
coefficients of the sub-blocks ${\bf K}$ and $\pmb{\Omega}_M$
approach the finite values for a flat disk (see
figure \ref{fig:sub_matrix_norms_full} below)
while the nonzero coefficients in the coupling matrix
${\bf C}_M$ go to zero, as shown in figure \ref{radius_variations}(c). However,
$C_{12}$ goes to zero much more quickly than $C_{21}$ (the data suggests that
$C_{12} \sim R_c^{-3}$ while $C_{21} \sim
R_c^{-1}$ as $R_c \to \infty$
which implies that $\pitch^{{\rm [centre]}} \to \pi/2$ as $R_c \to \infty$,
consistent with the behaviour observed in
figure \ref{radius_variations}(a), where the centres move towards
the right as $R_c$ is increased.

To interpret this result we note that for $\roll = \roll^{\rm [centre]} = \pi/2$ the
body-fitted normal ${\bf n}^{{\cal F}_b}$ (pointing in the direction
of $x_3^{{\cal F}_b}$) is oriented parallel to the
lab-frame $(x_1,x_2)$-plane, i.e. ${\bf n}^{{\cal F}_b} \cdot {\bf e}_3
= 0$. When the disk is in this orientation, the continous change
to $\yaw$ introduces a rigid body rotation
about $x_3$, while a change to $\pitch^{{\rm [centre]}}$ rotates the disk
about ${\bf n}^{{\cal F}_b}$ (recall again that the Tait-Bryan angles
describe rotations that have to be applied consecutively, in the
order yaw-pitch-roll). 

\begin{figure}
\begin{center}
  \includegraphics{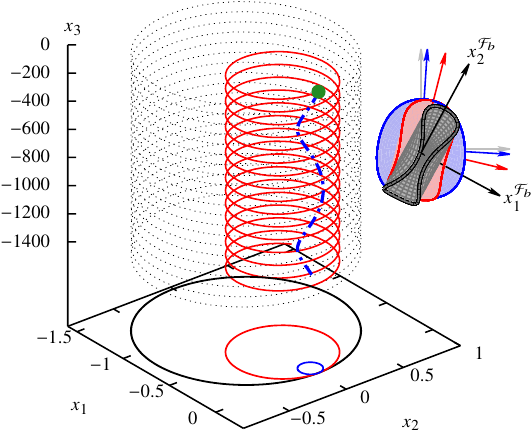}
  \caption{\label{radius_variations_for_fixed_point}The spiral
    trajectories of three disks with different curvatures, all released
    from ${\bf r}_M^{[0]}= {\bf 0}$ (green sphere), 
    with an initial inclination $\bigg( \pitch^{[0]},\roll^{[0]}\bigg) =
    \left(\pitch^{\rm [centre]}(R_c),\roll^{\rm [centre]} =
    \pi/2\right)$. The disk's inclination remains constant and it
    sediments along spiral trajectories. The inset shows the inclination of the disks
    relative to the direction of gravity. (Black: $R_c = 0.35$; red $R_c =
    0.5$; blue $R_c = 2$.) As $R_c \to \infty$ the spiral trajectories
    associated with this particular inclination degenerate to a
    straight vertical line and the disk sediments in an
    edge-on orientation.}
\end{center}
\end{figure}

The effect of changes to the disk's radius of curvature on its
orientation as it sediments
along the spiral trajectory associated with
$\bigg(\pitch^{{\rm [centre]}}(R_c),\roll^{{\rm
    [centre]}} = \pi/2 \bigg)$ is illustrated in
figure \ref{radius_variations_for_fixed_point}: for a tightly curved
disk, the body-fitted roll axis,
$x_2^{{\cal F}_b}$,  is strongly inclined away from the 
direction of gravity and the disk sediments along a wide spiral trajectory;
see the black disk and the associated black dotted line which shows
the trajectory of that disk's centre of mass in the lab frame.
As the radius of curvature increases and $\pitch^{{\rm
    [centre]}} \to \pi/2$, the disk rights itself and its roll axis approaches
the $x_3$-axis in the lab frame (red and blue disks). As it approaches this
configuration, the horizontal velocities go to zero while the vertical
velocity approaches the finite sedimentation speed of a flat disk in
its edge-on configuration. This leads to an increase in the pitch of the spiral
trajectories whose radius (in the $(x_1,x_2)$-plane) shrinks to zero. Thus
in the limit as $R_c \to \infty$,  a disk that sediments along the (nominally)
spiral trajectory associated with the centre in the
$(\pitch,\roll)$-phase plane sediments purely
vertically, and in an edge-on orientation.

\section{Discussion and Conclusions} 
We have studied the sedimention of U-shaped rigid disks at zero
Reynolds number and shown that, despite their non-chiral shape, they
tend to sediment along complex spiral paths whose chirality depends on
the disks' initial orientation. We showed this to be due to the fact that
the purely vertical sedimention of such disks in their upright U
orientation is stable (unstable) to perturbations about their roll (pitch)
axes, with the reverse being true when they sediment in an
upside-down U orientation. The instabilities cause the disks to
continuously reorient by undergoing alternating rolling- and
pitching-dominated rotations which are accompanied by a periodically
varying horizontal drift. During each roll/pitch cycle, the disk
performs a net rotation about the direction of gravity, which is generally an
irrational multiple of $\pi$. The
combination of these effects results in complex spiral trajectories,
similar to those we observed in our experimental study \citep{Miara2024}.

 The analysis was greatly facilitated by the fact that the numerical
solution of the 3D Stokes equations showed that wall effects (arising
from a finite computational domain or, in an experiment, the finite
size of the tank) only have a small effect on the pitching and rolling
instabilities that are responsible for the disk's behaviour. This
allowed us to analyse the disk's sedimentation in an unbounded fluid,
using resistance matrices computed in finite domains. While
this already led to a significant reduction in the computational effort
(compared to time-stepping the motion of the disk in the 3D fluid
domain), a further key simplification arose from the fact that for
sedimentation in an unbounded fluid, only two of the disk's six
rigid-body degrees of freedom (comprising the position of the disk's centre of
mass and its orientation, the latter described in terms of three Euler angles)
are genuine degrees of freedom. This allowed us to study the dynamics
of the disk's sedimentation in a two-dimensional phase space
comprising the Tait-Bryan pitch and roll angles. In this phase space,
the periodic rolling-pitching re-orientations appear as closed orbits.
They enclose a centre which corresponds to the special case in which the
disk sediments along a perfectly helical path while rotating at a
fixed rate about the direction of gravity.

 The behaviour of rigid U-shaped disks is therefore fundamentally
different from that of their flat counterparts which maintain
their inclination, and generally sediment with a constant
horizontal drift. The latter effect causes dilute clouds of
sedimenting flat disks to disperse horizontally; our findings imply
that clouds of U-shaped disks will not spread horizontally.

The difference in the behaviour of the two types of disks must, of course,
disappear when the curvature of the U-shaped disk is reduced to zero. We
showed that the rate at which U-shaped disks reorient
decreases with a reduction in their curvature, so that in the limit of
zero curvature the disk's initial inclination
changes infinitely slowly and thus remains constant.

We focused on the sedimentation of U-shaped disks because their shape
resembles that observed for elastic disks in preliminary experiments by
\cite{MiaraThesis} and also in recent computations by \cite{Yu23}
who found that, qualitatively, initially planar elastic
sheets tended to deform into such shapes as they sediment. 

The actual shape of a sedimenting elastic sheets is, of course, determined
by the fluid traction acting on them, and therefore has to
be obtained via the solution of a fully-coupled, and in general
time-dependent, fluid-structure interaction problem. However,
thin elastic sheets have an extensional (membrane/in-plane)
stiffness that is much larger than their bending stiffness. Therefore,
they tend to deform approximately isometrically, with little in-plane
stretching. This restricts the types of shapes that sedimenting
elastic sheets can adopt. Our results suggest that once the sheet has
deformed and reoriented itself into an upright U shape, it will
become subject to the (relatively slow) rotational instability about
the pitching axis. This implies that, counter to the conjecture by
\cite{Yu23}, the upright U-shape is not actually a stable configuration for 
elastic sheets.

\subsubsection*{Declaration of Interests}
The authors report no conflict of interest.

\subsubsection*{Acknowledgements}
Christian Vaquero-Stainer and Tymoteusz Miara
were funded by EPSRC DTP studentships. A.J. acknowledges funding by
EPSRC (Grant EP/T008725/1).

\bibliographystyle{jfm}
\bibliography{paper.bib}


\appendix
\section{Derivation of the evolution equations in an unbounded fluid\label{sec:angle_stuff}}

Figure \ref{fig:setup} shows how the disk's orientation
is described in terms of an intrinsic sequence of rotations by the
Tait-Bryan angles $\yaw, \pitch$ and $\roll$. Starting from a
reference orientation in which the disk's
principal axes are aligned with the Cartesian coordinates
$(x_1^{{\cal F}_{\rm lab}}, x_2^{{\cal F}_{\rm lab}},x_3^{{\cal F}_{\rm lab}})$ 
in the lab frame, ${\cal F}_{\rm lab}$, we apply rotations
about the three coordinate axes, but in subsequent frames, so that
\be
   \scalebox{1.5}{${\cal F}_{\rm lab}$}  = \scalebox{1.5}{${\cal F}_0$} \stackrel{
     \scalebox{0.8}{$\yaw$ about $x_3^{{\cal F}_{\rm lab}}$}}{ \xrightarrow{\hspace*{2.05cm}} }
   \scalebox{1.5}{${\cal F}_{1}$}  \stackrel{
     \scalebox{0.8}{$\pitch$ about $x_2^{{\cal F}_{\rm 1}}$}}{ \xrightarrow{\hspace*{2.05cm}} }
   \scalebox{1.5}{${\cal F}_{2}$}  \stackrel{
     \scalebox{0.8}{$\roll$ about $x_1^{{\cal F}_{2}}$}}{ \xrightarrow{\hspace*{2.05cm}} }
   \scalebox{1.5}{${\cal F}_{3}$} = \scalebox{1.5}{${\cal F}_{b}$}
   \label{eqn:sequence_of_rotations}
\ee
Given a vector ${\bf a}$ with components $a_i^{{\cal F}_{j}}$ in
frame ${\cal F}_{j}$, its components
in the subsequent frame are given by
\be
\bigg[
  a_{1}^{{\cal F}_{j+1}},
  a_{2}^{{\cal F}_{j+1}},
  a_{3}^{{\cal F}_{j+1}}
  \bigg]^T
= {\cal R}_{{\cal F}_{j}}^{{\cal F}_{j+1}} \ 
\bigg[
  a_{1}^{{\cal F}_{j}},
  a_{2}^{{\cal F}_{j}},
  a_{3}^{{\cal F}_{j}}
  \bigg]^T,
  \ee
where the matrices ${\cal R}_{{\cal F}_{j}}^{{\cal F}_{j+1}}$ are
standard orthogonal rotation matrices, e.g.
\be
{\cal R}_{{\cal F}_{0}}^{{\cal F}_{1}}
    =\left(
    \begin{array}{ccc}
      \cos(\yaw) & \sin(\yaw) & 0 \\
     -\sin(\yaw) & \cos(\yaw) & 0 \\
      0 & 0 & 1
      \end{array}
    \right)
    \mbox{\ \ \ with \ \ \ }
    {\cal R}_{{\cal F}_{1}}^{{\cal F}_{0}} =
    \bigg({\cal R}_{{\cal F}_{0}}^{{\cal F}_{1}}\bigg)^{-1} =
    \bigg({\cal R}_{{\cal F}_{0}}^{{\cal F}_{1}}\bigg)^{T},
\ee
and each of these matrices depend only on the single angle describing the
associated rotation. Thus translating the velocity of the disk's
centre of mass from ${\cal F}_{\rm lab}$ to
${\cal F}_{b}$ is achieved by
\be
\bigg[
  U_{1}^{{\cal F}_{\rm b}},
  U_{2}^{{\cal F}_{\rm b}},
  U_{3}^{{\cal F}_{\rm b}}
  \bigg]^T
= {\cal R}_{{\cal F}_{\rm lab}}^{{\cal F}_{\rm b}} \
\bigg[
  U_{1}^{{\cal F}_{\rm lab}},
  U_{2}^{{\cal F}_{\rm lab}},
  U_{3}^{{\cal F}_{\rm lab}}
  \bigg]^T,
\ee
where
\be
   {\cal R}_{{\cal F}_{\rm lab}}^{{\cal F}_{b}} =
   {\cal R}_{{\cal F}_{2}}^{{\cal F}_{b}}\
   {\cal R}_{{\cal F}_{1}}^{{\cal F}_{2}}\
   {\cal R}_{{\cal F}_{\rm lab}}^{{\cal F}_{1}}.
   \ee
Furthermore, the rate of change in the Tait-Bryan angles (which
perform rigid body rotations about coordinate axes aligned with subsequent
frames; see (\ref{eqn:sequence_of_rotations})) results in
an overall rate of rotation $\pmb{\omega}$ whose components in the body
frame are
\begin{eqnarray}
\bigg[
  \omega_{1}^{{\cal F}_{\rm b}},
  \omega_{2}^{{\cal F}_{\rm b}},
  \omega_{3}^{{\cal F}_{\rm b}}
  \bigg]^T
& = & 
{\cal R}_{{\cal F}_{2}}^{{\cal F}_{\rm b}} \ 
{\cal R}_{{\cal F}_{1}}^{{\cal F}_{2}} \
\bigg[ 0,0,\frac{{\rm d}\yaw}{{\rm d}t} \bigg]^T
+
{\cal R}_{{\cal F}_{2}}^{{\cal F}_{b}}\ 
\bigg[ \frac{{\rm d}\pitch}{{\rm d}t}, 0, 0 \bigg]^T
+
\bigg[ 0,\frac{{\rm d}\roll}{{\rm d}t}, 0 \bigg]^T \nonumber
\\
& = & {\cal S} \ \bigg[ 
  \frac{{\rm d}\pitch}{{\rm d}t},
  \frac{{\rm d}\roll}{{\rm d}t},
  \frac{{\rm d}\yaw}{{\rm d}t}
\bigg]^T,
\end{eqnarray}
where
\be
{\cal S} = 
\left(
\begin{array}{ccc}
  \cos(\roll) & 0 & -\cos(\pitch)\sin(\roll) \\
  0 & 1 & \sin(\pitch) \\
  \sin(\roll) & 0 &  \cos(\pitch)\cos(\roll) 
\end{array}
\right)
\ee
which is independent of $\yaw$.

This allows us to transform (\ref{eqn:fsi_governing_eqn_nondim2})
into quantities in the lab-frame of reference,
\begin{eqnarray}
\left(
\begin{array}{r}
     {\cal R}_{{\cal F}_{2}}^{{\cal F}_{b}}(\roll)\
     {\cal R}_{{\cal F}_{1}}^{{\cal F}_{2}}(\pitch)\
     {\cal R}_{{\cal F}_{\rm lab}}^{{\cal F}_{1}}(\yaw)
     \bigg[
       \frac{{\rm d}r_{M,1}^{{\cal F}_{\rm lab}}}{{\rm d}t},
       \frac{{\rm d}r_{M,2}^{{\cal F}_{\rm lab}}}{{\rm d}t},
       \frac{{\rm d}r_{M,3}^{{\cal F}_{\rm lab}}}{{\rm d}t}
       \bigg]^T \\
     {\cal S}(\pitch, \roll) \
     \bigg[
       \frac{{\rm d}\pitch}{{\rm d}t},
       \frac{{\rm d}\roll}{{\rm d}t},
       \frac{{\rm d}\yaw}{{\rm d}t}
       \bigg]^T
\end{array}
\right) = \nonumber \\
\label{big_system_in_lab_frame}
= \begin{pmatrix}
  {\bf K} & {\bf C}_M^T  \\
  {\bf C}_M & {\bf \Omega}_M
\end{pmatrix}^{-1}\left(
\begin{array}{r}
     {\cal R}_{{\cal F}_{2}}^{{\cal F}_{b}}(\roll)\
     {\cal R}_{{\cal F}_{1}}^{{\cal F}_{2}}(\pitch)\
     {\cal R}_{{\cal F}_{\rm lab}}^{{\cal F}_{1}}(\yaw)
     \bigg[
       F_1^{{\cal F}_{\rm lab}},
       F_2^{{\cal F}_{\rm lab}},
       F_3^{{\cal F}_{\rm lab}}
       \bigg]^T \\
     {\cal R}_{{\cal F}_{2}}^{{\cal F}_{b}}(\roll)\
     {\cal R}_{{\cal F}_{1}}^{{\cal F}_{2}}(\pitch)\
     {\cal R}_{{\cal F}_{\rm lab}}^{{\cal F}_{1}}(\yaw)
     \bigg[
       T_{{\rm pitch}}^{{\cal F}_{\rm lab}},
       T_{{\rm roll}}^{{\cal F}_{\rm lab}},
       T_{{\rm yaw}}^{{\cal F}_{\rm lab}}
       \bigg]^T
     \end{array}
\right),
\end{eqnarray}
which is easily re-written in the form (\ref{eqn:ode_system}). The
reason for keeping the equations in this form, and for spelling out
explicitly the dependence of the rotation matrices on the various
Tait-Bryan angles is that it
shows that the equations are not affected by
rigid body translations, $r_{M,i}^{{\cal F}_{\rm lab}}(t) \to
r_{M,i}^{{\cal F}_{\rm lab}}(t) + \delta r_{M,i}^{{\cal F}_{\rm lab}}$ for
arbitrary constants $\delta r_{M,i}^{{\cal F}_{\rm lab}}$, because none of
the coefficients depend on the position of the disk. Furthermore,
since for free sedimention we have  $\bigg[F_1^{{\cal F}_{\rm lab}},
       F_2^{{\cal F}_{\rm lab}},
       F_3^{{\cal F}_{\rm lab}}\bigg] = \bigg[0,0,-\pi h/{\cal L}\bigg]$
and $T_{\{{\rm roll,pitch,yaw}\}}^{{\cal F}_{\rm lab}}=0$, a change in the yaw angle by
an arbitrary constant $\delta \yaw$ so that
$\yaw(t) \to \yaw(t) + \delta \yaw$,
corresponding to a rigid body rotation about the vertical axis,
does not affect the right-hand side, while on the left-hand side
it simply subjects the trajectory to a rigid body rotation by that
same angle, i.e.
$\bigg[
  r_{M,1}^{{\cal F}_{b}}(t),
  r_{M,2}^{{\cal F}_{b}}(t),
  r_{M,3}^{{\cal F}_{b}}(t)
  \bigg]^T \to
{\cal R}_{\rm lab}^{1}(\delta \yaw) \
\bigg[
  r_{M,1}^{{\cal F}_{b}}(t),
  r_{M,2}^{{\cal F}_{b}}(t),
  r_{M,3}^{{\cal F}_{b}}(t)
  \bigg]^T$.
Thus, of the six initial conditions required for the solution of the
governing equations, four translate into trivial rigid body modes.
The shape of the trajectory and the rate at which the disk
reorients relative to the direction of gravity
while its centre of mass sediments along that path
depend only on the initial values for the pitch and roll angles.

In fact, these two angles are the system's only genuine degrees of
freedom: given the structure of
${\cal R}_{{\cal F}_{\rm lab}}^{{\cal F}_{1}}(\yaw)$, the
right-hand-side of (\ref{big_system_in_lab_frame}) depends only on 
$\pitch$ and  $\roll$, and it has non-zero
entries only in the first three rows. This implies that
$\pitch$ and  $\roll$ can be determined
independently from the fourth and fifth row of the ODE
system (\ref{big_system_in_lab_frame}) since these ODEs do not
involve $r_{M,i}^{{\cal F}_{\rm lab}}(t)$ and $\yaw$. This
leads to equations (\ref{pitch_and_roll_odes}). The evolution of the
yaw angle $\yaw(t)$ and the position of the centre of
mass, $r_{M,i}^{{\cal F}_{\rm lab}}(t)$,  can then be determined
a posteriori from the remaining four ODEs (equations (\ref{yaw_ode})
and (\ref{dr_Mdt_ode}) in \S \ref{sect:6x6}).

\section{The structure of the resistance matrix and the derived
  quantities \label{sec:grand_resistance_matrix}}
The U-shaped disks considered in this study have two
mutually-orthogonal planes of symmetry. 
\cite{HappelAndBrenner} show that for objects of this type
the translation, rotation and coupling tensors have the form
\begin{equation}
  \bm K =
  \begin{pmatrix}
    K_{11} & 0 & 0 \\
    0 & K_{22} & 0 \\
    0 & 0 & K_{33}
  \end{pmatrix},\quad
  (\bm \Omega_M) =
  \begin{pmatrix}
    \Omega^{(M)}_{11} & 0 & 0 \\
    0 & \Omega^{(M)}_{22} & 0 \\
    0 & 0 & \Omega^{(M)}_{33}
  \end{pmatrix}, \quad 
  \bm C_M =
  \begin{pmatrix} 
    0 & C^{(M)}_{12} & 0 \\
    C^{(M)}_{21} & 0 & 0 \\
    0 & 0 & 0
  \end{pmatrix}.  
  \label{eqn:tensors_two_planes_of_sym}
\end{equation}
The two diagonal blocks $\bm K$ and $\bm \Omega_M$ of the
resistance matrix are diagonal, just as in the case of a flat disk. The key
difference, which gives rise to non-trivial dynamics, arises from the
translation-rotation coupling described by $\bm C_M$. For flat disks
we have $\bm C_M = \bm 0$, allowing them to sediment along simple,
rotation-free trajectories. The fact that $C^{(M)}_{33}=0$
implies that our U-shaped disks can sediment along the body fitted
$x_3^{\mathcal F_b}$ directions without reorienting or
drifting sideways. This corresponds to the vertical sedimention of the
disk in its upright and upside-down U orientations. For all
other orientations the nonzero entries in $\bm C_M$ imply that a
translation in the $x_1^{\mathcal F_b}$ and $x_2^{\mathcal F_b}$
direction induce rotations about the $x_2^{\mathcal F_b}$ and $x_1^{\mathcal F_b}$
axes, respectively.

In terms of the nonzero entries in these three tensors, the
coefficients featuring in the ODEs (\ref{pitch_and_roll_odes}),
(\ref{yaw_ode}) and (\ref{dr_Mdt_ode}) in \S \ref{sect:6x6} are then given by

\be
\mathbb{A} = \mathit{C_{21}} \mathit{K_{33}} \Omega_{33} 
\left(\mathit{K_{11}}  \Omega_{22} -\mathit{C_{12}}^{2} \right), 
\ee

\be
\mathbb{B} = \mathit{K_{33}} \Omega_{33}
\left(
\mathit{C_{12}}^{2} \mathit{C_{21}}  +
\mathit{C_{12}} \,\mathit{C_{21}}^{2} 
-\mathit{C_{12}} \mathit{K_{22}}  \Omega_{11} 
-\mathit{C_{21}} \mathit{K_{11}}  \Omega_{22}
\right),
\ee

\be
\mathbb{D} = \mathit{C_{21}} \mathit{K_{33}} \Omega_{33}
\left(\mathit{C_{12}}^{2}-\mathit{K_{11}}  \Omega_{22}\right),
\ee

\be
\mathbb{E} = 
\left(\mathit{C_{21}}^{2} \mathit{K_{33}} -\mathit{K_{22}} \mathit{K_{33}}
\Omega_{11} \right) \left(\mathit{C_{12}}^{2} \Omega_{33} -\mathit{K_{11}}
\Omega_{22} \Omega_{33} \right),
\ee

\be
\mathbb{F} = 
\left(\mathit{C_{21}}^{2}-\mathit{K_{22}} \Omega_{11} \right)
\left(\left(\mathit{K_{33}} -\mathit{K_{11}} \right) \Omega_{22}
+\mathit{C_{12}}^{2}\right),
\ee

\be
\mathbb{G} = 
\left(\mathit{C_{21}}^{2}-\mathit{K_{22}} \Omega_{11} \right)
\left(\left(\mathit{K_{33}} -\mathit{K_{11}} \right) \Omega_{22}
+\mathit{C_{12}}^{2}\right),
\ee

\be
\mathbb{H} =
\mathit{K_{33}} \left(\left(\left(-\mathit{K_{11}} +\mathit{K_{22}} \right)
\Omega_{11} -\mathit{C_{21}}^{2}\right) \Omega_{22} +\mathit{C_{12}}^{2} \Omega_{11}
\right),
\ee

\be
\mathbb{I} = 
\mathit{K_{33}} \left(\left(\left(-\mathit{K_{11}} +\mathit{K_{22}} \right)
\Omega_{22} +\mathit{C_{12}}^{2}\right) \Omega_{11} -\mathit{C_{21}}^{2} \Omega_{22}
\right),
\ee

\be
\mathbb{J} = 
\mathit{K_{33}} \Omega_{11} \left(\mathit{C_{12}}^{2}-\mathit{K_{11}} \Omega_{22}
\right),
\ee

\be
\mathbb{K} =
\mathit{K_{33}} \left(\mathit{C_{21}}^{2}-\mathit{K_{22}} \Omega_{11} \right) \left(\mathit{C_{12}}^{2}-\mathit{K_{11}} \Omega_{22} \right).
\ee

\section{\label{sec:wall_effects}The importance of wall effects}
The analysis presented in this paper relies heavily on the observation
that wall effects only have a modest effect on the rolling and
pitching behaviour of the disk. This allowed us to compute the 
resistance matrix for a disk sedimenting in an unbounded fluid
by performing computations in a large but finite computational
domain. The subsequent analysis was then performed by by analysing
the solutions of a small system of ODEs.

\begin{figure}
  \begin{center}
  \includegraphics{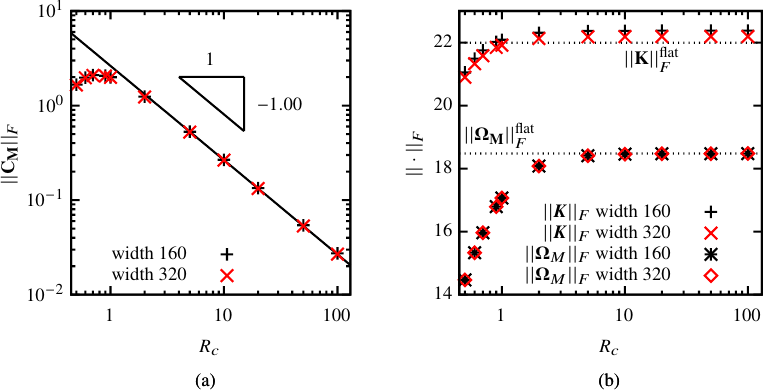}
  \caption{\label{fig:sub_matrix_norms_full}The Frobenius norms of the 
    sub-matrices $\bm K, \bm \Omega_M$ and $\bm C_M$, which comprise
    the resistance matrix, shown as a function of the radius of 
    curvature of the disk and for box sizes $L=160$ (black `+') and $L=320$ (red `$\times$').
    (a) Log-log plot of the norm of the coupling tensor $\bm C_M$;
    the solid line shows a power-law fit to the final
    five data points, with the fitted exponent shown via the slope-triangle;
    (b) Semi-log plot of the linear and rotational drag tensors $\bm K$ and
    $\bm \Omega_M$; the dotted lines indicate the analytic values for a
    flat circular disk in an unbounded fluid.
  }    
\end{center}
\end{figure}

Given that in Stokes flow boundary effects are known to act over large
distances, we briefly revisit the importance of the domain size (which
features both in computations and in experiments) on our
results. We showed in \S \ref{sec:radius_effects} that the (in
general) chiral sedimention of U-shaped disks is due the non-zero
off-diagonal blocks ${\bf C}_M$ in the resistance matrix. We
reconciled the behaviour of U-shaped disks with the flat counterpart
by showing that these blocks to go zero as the disk's radius of
curvature, $R_c$, increases. Figure \ref{fig:sub_matrix_norms_full}(a) shows
that the dependence  of the off diagonal block ${\bf C}_M$ on
the disk's radius of curvature is not visibly affected by an increase
in the size of the computational domain. Conversely, the
corresponding plots for the Frobenius norms of the two diagonal
blocks ${\bf K}_M$  and $\pmb{\Omega}_M$ in figure
\ref{fig:sub_matrix_norms_full}(b) show that, as the disk's radius of
curvature is increased, both norms approach constant values which are slightly
above the theoretical values of $16 \sqrt{17}/3$ and $32 \sqrt{3}/3$
for a flat disk sedimenting in an unbounded fluid. The data computed
in a domain of twice the (linear) size can be seen
to be closer to the theoretical value for an unbounded fluid which
is approached from above, indicating that the finite-size effects
increase the translational and rotational drag \citep{Brenner62, Caswell72}. 

\begin{figure}
 \begin{center}
  \includegraphics{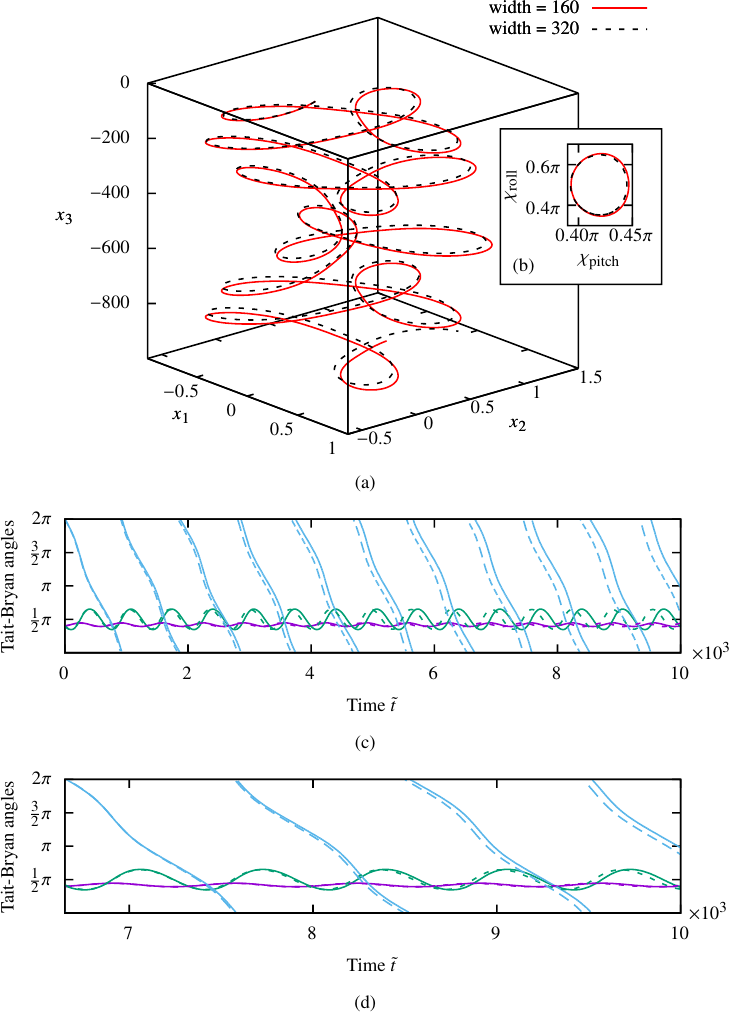}
  \caption{Trajectories for a disk with $R_c=0.5$, released with initial conditions
    $\pitch^{[0]} = 0.4\pi, \roll^{[0]} = 0.4\pi$. Data was obtained
    with resistance matrices computed in domains of different sizes:
    $L = 160$ (solid) and $L = 320$ (dashed).
    (a) Trajectories of the centres of mass;
    (b) phase-space trajectory of the disk orientation;
    (c) time trace of the Tait-Bryan angles;
    (d) zoomed time trace of the Tait-Bryan angles,
    with a time offset applied
    to the $L=320$ (dashed) trace to align the two
    trajectories at $\tilde{t} = 6.65$.
    \label{fig:box_width_comparison_trajectories}
    }
 \end{center}
\end{figure}

To assess how this affects the disks's behaviour
figure \ref{fig:box_width_comparison_trajectories} shows representative
plots of (a) the trajectory of the disk's centre of mass, 
(b) the disk's closed orbits in the $(\pitch,\roll)$-phase plane, and
(c,d) the time evolution of the three Tait-Bryan angles. Solid and dashed lines
represent the predictions obtained using the resistance matrix
computed in domains of size ${\tt L} = 160$ and ${\tt L} = 320$,
respectively. As anticipated from all the results presented so far,
the $(\pitch,\roll)$-phase planes obtained from
the two matrices are virtually identical.
The slightly increased drag in the smaller domain results in
small changes to the disk's predicted
velocity, causing the two centre-of-mass trajectories shown in
figure \ref{fig:box_width_comparison_trajectories}(a) to drift apart
while the disk undergoes its periodic pitch-roll cycle.
The time trace of the Tait-Bryan angles shows that the 
resistance matrix computed in the larger domain leads to a slight
decrease in the period of the pitch-roll cycle which again causes the
time-traces to drift apart. However this has no effect on the qualitative
behaviour within each roll-pitch cycle. This is illustrated by the
zoomed-in plot shown in figure \ref{fig:box_width_comparison_trajectories}(d),
where the dash-dotted line was obtained by adding a suitable time
offset to the dashed line in
figure \ref{fig:box_width_comparison_trajectories}(c),
so that the two time-traces are (re-)aligned at the beginning of that
period. Following this re-aligment of the curves,
the two time-traces are nearly indistinguishable again.

\section{Validation}

In order to construct a suitable validation case for our numerical scheme,
we considered the special case of a flat, circular disk moving in an unbounded
and otherwise quiescent fluid; for this case we have a closed-form
exact solution $({\bf u}_{\rm exact},p_{\rm exact})$
\citep{HappelAndBrenner}, for which the force and torque on the disk are given by
\begin{align}
  \bm F_{\rm exact} = \frac{16}{3}
  \begin{pmatrix}
    2 & 0 & 0\\
    0 & 2 & 0 \\
    0 & 0 & 3
  \end{pmatrix}
  \cdot \bm U_M, \quad\quad
  \bm T_{\rm exact} = \frac{32}{3}
  \begin{pmatrix}
    1 & 0 & 0\\
    0 & 1 & 0 \\
    0 & 0 & 1
  \end{pmatrix} \cdot \bm \omega.
  \label{eqn:analytic_force_and_torque_flat_disk}
\end{align}

\begin{figure}
   \begin{center}
  \includegraphics{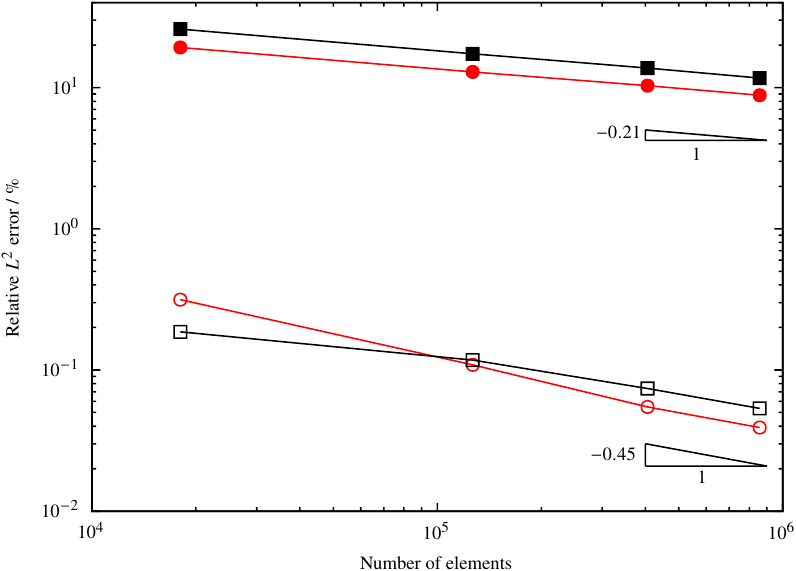}
  \caption{Relative error (in percent) of the computed drag (circles)
  and torque (squares)
  as a function of the mesh refinement, for a validation case in
  which a flat circular disk moves with
  $\bm U_M = \{-0.7, -0.3, 0.6\}, \bm \omega =\{-3.1, 0.1, 0.9\}$.
  Filled symbols are for classical Taylor-Hood elements; hollow
  symbols are for the PDE-constrained solution where the velocity and
  pressure fields are augmented with appropriate singular
  functions, as described in \S \ref{sec:augmented_stokes}.
  \label{fig:fe_vs_aug_validation}}
 \end{center}
\end{figure}

We chose an arbitrary rigid body motion of the disk,
$\bm U_M = \{-0.7, -0.3, 0.6\}, \bm \omega =\{-3.1, 0.1, 0.9\}$, which
we imposed via the boundary conditions (\ref{eqn:no_slip}) on the disk
surface. On the outer boundaries of the computational domain 
(a cuboid of size $1.55\times 1.55 \times 0.55$) we we applied the
analytical velocity field as a Dirichlet conditions,
except for the top surface where we imposed
the analytical traction $\bm \tau_{\rm exact} \cdot \bm n$ as a Neumann condition. We then
uniformly refined the mesh several times, and for each mesh computed the
numerical solution using both classical Taylor-Hood elements and our augmented
finite elements. The results of this convergence study are shown in
figure \ref{fig:fe_vs_aug_validation}, where we show the relative error between the
analytical and computed values for drag and torque, obtained with both
methods, as a function of the number of elements in the mesh. The
figure demonstates that the augmentation by singular functions
described in \S \ref{sec:augmented_stokes} not only outperforms the
non-augmented elements by two orders of magnitude, but converges at a rate of
more than twice the asymptotic order under mesh refinement.

\end{document}